\documentclass[english,aps,prb,showpacs,twocolumn]{revtex4-2}
\usepackage[T1]{fontenc}
\usepackage[latin9]{inputenc}
\usepackage{babel}
\usepackage{graphicx}
\usepackage{bm}
\usepackage{bbm}
\usepackage{array}
\usepackage{amsmath}
\usepackage{amssymb}
\usepackage{dcolumn}
\usepackage{epstopdf}
\usepackage{ulem}
\usepackage{bbold}
\usepackage{tikz}
\usepackage[margin=15mm]{geometry}
\usepackage[colorlinks=true,linkcolor=blue,citecolor=blue,urlcolor=blue]{hyperref}
\usetikzlibrary{shadings}

\usepackage{color}

\begin{document}

\begin{abstract}

The effect of inertial spin dynamics is compared between ferromagnetic, antiferromagnetic and ferrimagnetic systems. The linear response to an oscillating external magnetic field is calculated within the framework of the inertial Landau--Lifshitz--Gilbert equation using analytical theory and computer simulations. Precession and nutation resonance peaks are identified, and it is demonstrated that the precession frequencies are reduced by the spin inertia, while the lifetime of the excitations is enhanced. The interplay between precession and nutation is found to be the most prominent in antiferromagnets, where the timescale of the exchange-driven sublattice dynamics is comparable to inertial relaxation times. Consequently, antiferromagnetic resonance techniques should be better suited for the search for intrinsic inertial spin dynamics on ultrafast timescales than ferromagnetic resonance.

\end{abstract}

\title{Nutation in antiferromagnetic resonance}

\author{Ritwik Mondal}
\email[]{ritwik.mondal@uni-konstanz.de}
\affiliation{ Fachbereich Physik, Universit\"at Konstanz, DE-78457 Konstanz, Germany}
\author{Sebastian Gro{\ss}enbach}
\affiliation{ Fachbereich Physik, Universit\"at Konstanz, DE-78457 Konstanz, Germany}
\author{Levente R\'ozsa}
\affiliation{ Fachbereich Physik, Universit\"at Konstanz, DE-78457 Konstanz, Germany}
\author{Ulrich Nowak}
\affiliation{ Fachbereich Physik, Universit\"at Konstanz, DE-78457 Konstanz, Germany}
\date{\today}


\maketitle

{ \section{Introduction}}
Deterministic spin switching at ultrashort timescales builds the fundament for future spin-based memory technology~\cite{Stanciu2007,Vahaplar2009,Radu2011,vahaplar12,Hass2013}. 
{ At femtosecond timescales inertial switching becomes particularly relevant}, where the reversal is achieved with a linear momentum gained by the interaction of an ultrashort pulse and spin inertia~\cite{Kimel2009,Wienholdt2012PRL}. 
The understanding of magnetic inertia has been pursued along two different directions so far.

On the one hand, spin dynamics in antiferromagnets (AFMs) and ferrimagnets (FiMs) has successfully been described by the Landau--Lifshitz--Gilbert (LLG) equation~\cite{landau35,Gilbert1955,Gilbert2004} for two sublattices coupled by the exchange interaction. The exchange energy created by tilting the sublattice magnetization directions away from the antiferromagnetic orientation is dynamically transformed into anisotropy energy by collectively rotating the sublattices away from the easy magnetic direction~\cite{Rozsa}, analogously to the transition between kinetic and potential energy terms in a harmonic oscillator. While the LLG equation for the two sublattices is of first order in time, this effect gives rise to an effectively inertial second-order differential equation for the order parameter in AFMs~\cite{Gomonai,Hals}. The interaction between exchange and anisotropy degrees of freedom causes an exchange enhancement of AFM resonance frequencies and linewidths~\cite{Gurevich}.

On the other hand, an intrinsic inertia also arises in magnetic systems, 
if it is assumed that the directions of spin angular and magnetic moments become separated in the ultrafast dynamical regime~\cite{Ciornei2011,Wegrowe2012}. The inertia gives rise to spin nutation, a rotation of the magnetization around the angular momentum direction~\cite{Bottcher2012}, caused by the energy transfer between magnetic kinetic and potential energy terms. The emergence of spin inertia has been explained based on an extension of the breathing Fermi surface model \cite{Fahnle2011,Fahnle2011JPCM}, calculated from a $s-d$ like interaction between the magnetization density and electron spin~\cite{Bhattacharjee2012} and derived from a fundamental relativistic Dirac theory~\cite{Mondal2017Nutation,Mondal2018JPCM}. Magnetic inertia can be associated with a torque term containing a second-order time derivative of the magnetic moment appearing in the inertial LLG (ILLG) dynamical equation. 
The characteristic inertial relaxation time, { using its definition in Eq.~\eqref{LLGN} below,} is expected to range from 1~fs~\cite{Ciornei2011,Bhattacharjee2012,Li2015,Thonig2017} to a few hundred fs~\cite{neeraj2019experimental}.

Linear-response theory predicted the emergence of a nutation resonance besides the conventional precession resonance in ferromagnets (FMs)~\cite{Olive2012,Olive2015,cherkasskii2020nutation}, providing a possible way of detecting inertial dynamics by applying oscillating external fields. An indirect evidence of the inertial dynamics was found in NiFe and Co samples~\cite{Li2015} by following the field dependence of the ferromagnetic precession resonance (FMR) peaks. The experimental observation of the nutation resonance has only been achieved very recently in NiFe and CoFeB using intense terahertz magnetic field transients~\cite{neeraj2019experimental}.

While the notion of inertial dynamics has been applied both in the context of the LLG equation { for} AFMs as well as in the ILLG equation for FMs, the linear response of these two examples is fundamentally different. 
While in both cases a pair of resonances is found in contrast to the single FMR peak, the excitation frequencies in an AFM are degenerate in the absence of a static external field, while they differ by several orders of magnitude in the ILLG equation. { The} effective damping parameter { of the precession}, defined as the half-width of the peak at half-maximum, is considerably higher in AFMs than in FMs, where it corresponds to the Gilbert damping. In contrast, it was demonstrated that the effective damping decreases in the ILLG equation applied to FMs~\cite{Olive2015}, particularly at the nutation resonance~\cite{Makhfudz2020}. 
However, the ILLG has { not been applied to AFMs so far}.

Here, we explore the effects of the ILLG equation in two-sublattice AFMs and FiMs using linear-response theory and computer simulations. It is shown that a pair of nutation resonance peaks emerges, and that the inertial relaxation time influences the precessional resonance significantly stronger in AFMs than in FMs due to the exchange coupling between the sublattices. The effective damping parameter is found { to} decrease in AFMs, reaching considerably lower values than the Gilbert damping at the nutation peak, thereby enhancing the lifetime of these excitations. The inertial effects in FiMs are found to interpolate between those in AFMs and FMs.        

{ \section{ Methods}}

As derived in earlier works~\cite{Ciornei2011,Mondal2017Nutation,Mondal2018JPCM}, the ILLG equation reads
\begin{align}
\label{LLGN}
    \dot{\bm{M}_{i}} & = - \gamma_{i} \bm{M}_{i}\times \bm{H}_{i} + \frac{\alpha_{i}}{M_{i0}} \bm{M}_{i}\times\dot{ \bm{M}_{i}} + \frac{\eta_i}{M_{i0}}\bm{M}_{i}\times\ddot{ \bm{M}_{i}}\,,
\end{align}
generalized here to multiple sublattices indexed by $i$. The first, second and third terms in Eq.~\eqref{LLGN} describe spin precession with gyromagnetic ratio $\gamma_{i}$, transverse relaxation with Gilbert damping $\alpha_{i}$, and inertial dynamics with relaxation time $\eta_{i}$. Note that an alternative notation for the inertial term with $\eta_{i}=\alpha_{i}\tau_{i}$ is also used in the literature~\cite{Ciornei2011,Li2015,neeraj2019experimental}; where comparison with earlier works is mentioned in the following, the relaxation time is converted to the formulation of Eq.~\eqref{LLGN}. {{The equation of motion was treated analytically as described in the following sections, and also solved numerically using an algorithm presented in detail in Appendix~\ref{appendixA}.}}

{ \section{ Inertial effects in ferromagnets}}

\begin{figure*}
\centering
\includegraphics[width = 1.8\columnwidth]{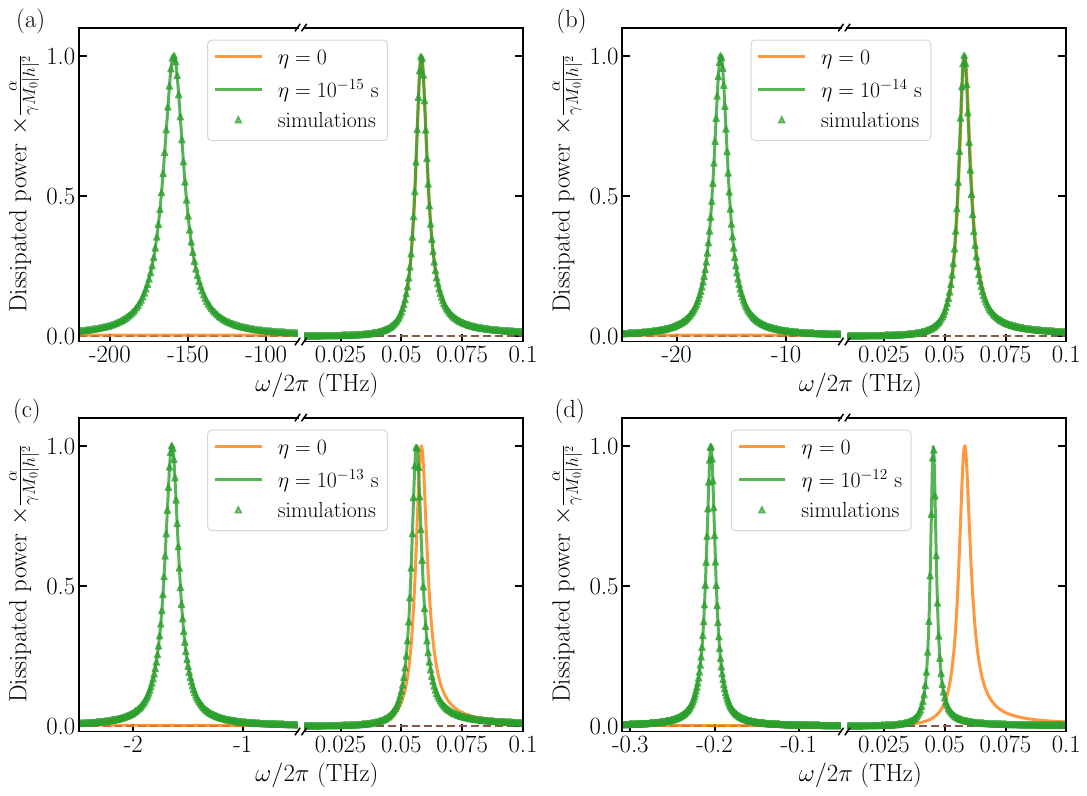}
\caption{ 
The rate of energy dissipation in the ferromagnet as a function of frequency for several values of the inertial relaxation time, (a) $\eta = 1$~fs, (b) $\eta = 10$~fs, (c) $\eta = 100$~fs, and (d) $\eta = 1$~ps. The lines denote the results of the analytical calculations and the symbols of the atomistic simulations for a single macrospin. All curves are compared to the analytical expression obtained without the inertial term. The other parameters are $\gamma = 1.76\times 10^{11}$~T$^{-1}$s$^{-1}$, $M_{0} = 2 \mu_{\textrm{B}}$, $ H_0 = 1$~T, $K = 10^{-23}$~J, $\alpha = 0.05$, and $\vert h\vert = 0.001$~T.} 
\label{dissipation_ferro}
\end{figure*}

First, we summarize the effects of the inertial term on FM resonance. The FM is described by the free energy $\mathcal{F}\left(\bm{M}\right) = - H_0 M_{z} - K M_{z}^2/M_{0}^{2} $, modeling a single sublattice where spatial modulations of the magnetization are neglected. $M_0$ is the magnitude of the magnetic moment, $H_0$ is the applied external field and $K$ is the uniaxial anisotropy energy, also considered to include demagnetization effects in the form of a shape anisotropy. The effective field can be written as $\bm{H}  =-\partial\mathcal{F}/\partial\bm{M}= (H_{\rm 0} + 2KM_{z}/M^{2}_{0})\hat{\bm{e}}_z$, and the magnetic moment is oriented along the $z$ direction in equilibrium. 

The linear response to a small transversal external field component $\bm{h}(t)$ is calculated considering $\bm{M}  = M_{0} \hat{\bm{e}}_z + \bm{m}(t)$ and expanding Eq.~\eqref{LLGN} up to first order in $\bm{h}(t)$ and $\bm{m}(t)$. The exciting field is assumed to be circularly polarized, $h_{\pm}=h_{x}\pm\textrm{i}h_{y}=h\textrm{e}^{\pm\textrm{i}\omega t}$, with a similar time dependence for the response, $m_{\pm}=m_{x}\pm\textrm{i}m_{y}=m\textrm{e}^{\pm\textrm{i}\omega t}$. The calculated susceptibility reads (see { Appendix~\ref{appendixB}} for details)
\begin{align}
m_{\pm}=\chi_{\pm}h_{\pm} &  = \frac{\gamma M_0}{\Omega_{0} - \omega   - \eta \omega^2 \pm \textrm{i} \alpha \omega }h_{\pm}\,, 
\label{suscFM}
\end{align}
with $\Omega_{0} = \gamma \left( H_{0} M_0+ 2K\right) / M_0$. It is found that the Gilbert damping is associated with the imaginary part of the susceptibility, while the inertial term contributes to the real part of the susceptibility, which is consistent with the previous calculation { in Ref.}~\cite{Mondal2017Nutation}. The dissipated power is calculated as $P=\dot{\bm{m}} \cdot \bm{h}=\omega {\tt Im}(\chi_+)\left|h\right|^{2}$. {{We note that a linearly polarized exciting field can be described as a linear combination of circularly polarized fields with $\omega$ and $-\omega$ frequencies.}}

The dissipated power with and without the inertial term is shown in Fig.~\ref{dissipation_ferro}. { The data points denoted by symbols in Fig.~\ref{dissipation_ferro} denote the results of the atomistic spin simulations (see Appendix~\ref{appendixA} for details).}
The relaxation time is chosen to {{range from $\eta = 10^{-15}$~s to $\eta = 10^{-12}$~s. This covers the fs timescales described in Refs.~\cite{Li2015,Thonig2017,Bhattacharjee2012} and the values of around 300~fs in Ref.~\cite{neeraj2019experimental}.}} 
It can be observed that the inertial dynamics reduces the precession resonance frequency. The resonance peak position is well approximated 
as $\omega_{\textrm{p}}=\left(\sqrt{1+4\beta_{\textrm{FM}}}-1\right)/\left(2\eta\right)\approx \Omega_{0}\left(1-\beta_{\textrm{FM}}\right)$, with $\beta_{\textrm{FM}}=\eta\Omega_{0}$. The associated shift in the resonance field $H_{\textrm{p}}$ was investigated in Ref.~\cite{Li2015}. However, note that the relative value of this shift is very low since $\beta_{\textrm{FM}}\ll 1$, meaning that it can only be observed if $\Omega_{0}$ is shifted to high values, for example by a strong external field $H_{0}$.

The most profound effect of the inertial dynamics is the emergence of a second resonance peak, associated with the spin nutation. Its frequency is approximately $\omega_{\textrm{n}}=-\left(\sqrt{1+4{ \beta_{\textrm{FM}}}}+1\right)/\left(2\eta\right)\approx -1/\eta-\Omega_{0}\left(1-{ \beta_{\textrm{FM}}}\right)$. 
Similarly to the precession frequency, the subleading corrections { $\beta_{\rm FM}\Omega_0$} are small. The negative sign of the frequency implies an opposite rotational sense~\cite{Kikuchi}: while the precession is excited by a circularly polarized field rotating counterclockwise, the nutation resonance reveals an opposite polarization.
 
The effective damping parameter is defined as the ratio of the imaginary and the real parts of the frequency where Eq.~\eqref{suscFM} has a node, and is approximately expressed as $\alpha_{\textrm{eff,p}}=\alpha_{\textrm{eff,n}}\approx\alpha\left(1-2{ \beta_{\textrm{FM}}}\right)$, see { Appendix~\ref{appendixB}} for the derivation. Since the imaginary part characterizes the half-width of the resonance peak at half maximum, the latter suggests that the linewidth of FMR decreases due to the inertia, in agreement with the numerical results in Ref.~\cite{Olive2015}. The relative value of the reduction is once again governed by the factor ${\beta_{\textrm{FM}}}$.

{ \section{Inertial effects in antiferromagnets and ferrimagnets}}

Next, we consider AFMs and FiMs with two sublattices $A$ and $B$. 
Assuming once again homogeneous sublattice magnetizations, the free energy is expressed as
\begin{align}
    & \mathcal{F}\left(\bm{M}_{A}, \bm{M}_{B}\right)   =  - H_0\left( M_{Az}+ M_{Bz}\right) \nonumber\\
    & - \frac{K_A}{M^{2}_{A0}} M^2_{Az} - \frac{K_B}{M^{2}_{B0}} M^2_{Bz} + \frac{J}{M_{A0}M_{B0}}  \bm{M}_A\cdot \bm{M}_B\,, 
     \label{Free-energy}
\end{align}
with the external field applied along the $z$ direction, $\bm{H}_0 = H_0\hat{\bm{e}}_z$, uniaxial easy-axis anisotropy constants $K_{A},K_{B}$ and intersublattice exchange coupling $J$. From the free energy, the associated fields entering the sublattice ILLG equations~\eqref{LLGN} can be determined using $\bm{H}_{A/B} = - \partial \mathcal{F}\left(\bm{M}_A, \bm{M}_B\right)/\partial \bm{M}_{A/B}= H_0\hat{\bm{e}}_z + 2K_{A/B}M_{A/Bz}/M^{2}_{A/B0}\hat{\bm{e}}_z-J\bm{M}_{B/A}/\left(M_{A0}M_{B0}\right)$. In equilibrium, the sublattice magnetizations are aligned antiparallel along the $z$ direction. Linear response to the transverse homogeneous external field $\bm{h}_{A}(t)=\bm{h}_{B}(t)$ may be calculated similarly to the FM case, using the expansions $\bm{M}_A(\bm{r},t) = M_{A0}\hat{\bm{e}}_z + {\bm{m}}_A(t)$ and $\bm{M}_B(\bm{r},t) = - M_{B0}\hat{\bm{e}}_z +{\bm{m}}_B(t)$.

The two-sublattice susceptibility tensor is expressed as follows (see { Appendix~\ref{appendixC} for details}):
\begin{widetext}
\begin{align}
    \begin{pmatrix}m_{A\pm} \\ m_{B\pm}\end{pmatrix}=\chi_{\pm}^{AB}\begin{pmatrix}h_{A\pm} \\ h_{B\pm}\end{pmatrix} & =  \frac{1}{\Delta_\pm}\begin{pmatrix}
     \frac{1}{ \gamma_B M_{B0}}\left(\Omega_B \pm \textrm{i}\omega\alpha_{B} -\eta_{B}\omega^2 +\omega  \right) &  -\frac{1}{M_{A0}M_{B0}}J\\
     - \frac{1}{M_{A0}M_{B0}}J &   \frac{1}{ \gamma_A M_{A0}}\left(\Omega_A \pm \textrm{i}\omega\alpha_{A} -\eta_{A}\omega^2-\omega  \right)
    \end{pmatrix}\begin{pmatrix}h_{A\pm} \\ h_{B\pm}\end{pmatrix}\,,\label{suscAFM}
\end{align}
Here we use the definitions 
$\Delta_\pm = \left(\gamma_A M_{A0} \gamma_B M_{B0}\right)^{-1}\left(\Omega_A \pm \textrm{i}\omega\alpha_{A} -\eta_{A}\omega^2-\omega  \right)\left(\Omega_B \pm \textrm{i}\omega\alpha_{B} -\eta_{B}\omega^2 +\omega  \right) - J^2/\left(M_{A0}^2 M_{B0}^2\right)$ as well as $\Omega_A =  \gamma_A/M_{A0}( J  + 2K_A +  H_0 M_{A0})$ and $\Omega_B = \gamma_B/M_{B0}( J + 2K_B - H_0 M_{B0})$.
\end{widetext}
\begin{figure*}
\centering
\includegraphics[width = 0.8\columnwidth]{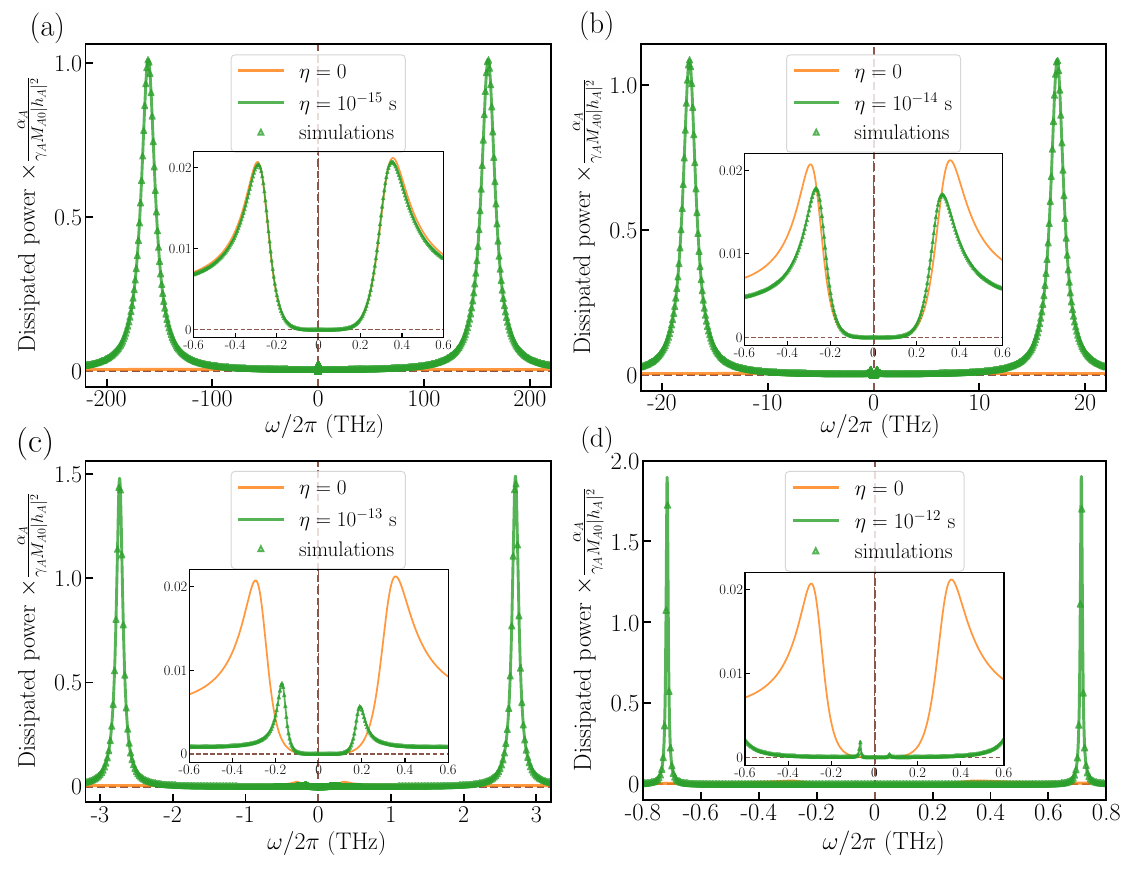}
\caption{ 
The rate of energy dissipation for the antiferromagnet as a function of frequency for several values of the inertial relaxation time $\eta_{A} = \eta_{B} = \eta$, (a) $\eta = 1$~fs, (b) $\eta = 10$~fs, (c) $\eta = 100$~fs and (d) $\eta = 1$~ps. The lines denote the results of the analytical calculations and the symbols of the atomistic spin simulations for two coupled macrospins. All curves are compared to the analytical expression obtained without the inertial term. The other parameters are $M_{A0} = M_{B0} = 2 \mu_{\textrm{B}}$, $ \gamma_A = \gamma_B  = 1.76 \times 10^{11}$~T$^{-1}$s$^{-1}$, $\alpha_{A} = \alpha_{B} = 0.05$, $K_A = K_B = 10^{-23}$~J, $J = 10^{-21}$~J, $ H_0 = 1$~T, and $\vert h_A\vert = \vert h_B \vert = 0.001$~T. The insets show the precession resonances on a smaller frequency and power scale. 
} 
\label{dissipation_antiferro}
\end{figure*}

To compare with FMR, we compute the dissipated power for AFMR, $P=\dot{\bm{m}}_{A} \cdot \bm{h}_{A}+\dot{\bm{m}}_{B} \cdot \bm{h}_{B}$, with the explicit formula given in { Appendix~\ref{appendixC}}.
The result is shown in Fig.~\ref{dissipation_antiferro}, using the same parameters for both sublattices as for the FM in Fig.~\ref{dissipation_ferro}. 
The insets of Fig.~\ref{dissipation_antiferro} show that without the inertial term the AFM precession resonance peaks are suppressed with respect to the { FM} one by a factor of about $J/\left(2K\right)=50$. This is caused by the fact that the magnetization in the two sublattices rotates { around} the equilibrium direction { with a phase shift of $\pi$}, meaning that the homogeneous exciting field only couples to the difference of the sublattice precession amplitudes~\cite{Gurevich} in the dissipated power. Also, the inertial term shifts the precession resonance peaks to lower frequencies considerably stronger than in the FM, and further reduces their magnitude. At higher frequency, two additional nutation resonance peaks can be observed.  Remarkably, their height is significantly larger than that of the precession resonances, even exceeding the intensity of the FMR peaks (cf. Fig.~\ref{dissipation_ferro} where the same normalization was used). The latter suggests that probing the AFM nutation resonance peak is experimentally more suitable than in the FM case. Most of these effects can be explained by the fact that the precession and nutation resonance frequencies lie much closer in AFMs than in FMs, as will be discussed in detail below.

To obtain the AFM resonance frequencies, we calculate the nodes of the susceptibility tensor in Eq.~\eqref{suscAFM}, obtaining
\begin{align}
    \Delta_{\pm}=a_{\pm}\omega^4 + b_{\pm}\omega^3 + c_{\pm}\omega^2 + d_{\pm}\omega +e_{\pm} = 0\,.
    \label{generic-equation}
\end{align}
with the following definitions: 
\begin{align}
    a_{\pm} &  = \eta_{A}\eta_{B}\,,\\
    b_{\pm} & = \mp \textrm{i} \left(\alpha_{A}\eta_{B}+\alpha_{B}\eta_{A}\right)- \left( \eta_{A} -\eta_{B} \right)\,,\\
    c_{\pm} & = -1 \pm \textrm{i}(\alpha_{A} - \alpha_{B}) - (\Omega_A\eta_{B} + \Omega_{B} \eta_{A}) \nonumber\\
    & - \alpha_{A}\alpha_{B}\,,\\
    d_{\pm} & =  \left(\Omega_A - \Omega_B\right) \pm \textrm{i}\left(\alpha_{B}\Omega_A + \alpha_{A}\Omega_B\right)\,,\\
    e_{\pm} & = -\frac{ \gamma_A}{M_{A0}}\frac{ \gamma_B}{M_{B0}} J^2  + \Omega_A\Omega_B\,.
\end{align}
Note that inertial effects enter via $a, b$, and $c$, terms which are of higher order in frequency. Setting the inertial relaxation times to zero, we obtain a second-order equation that results in well-known antiferromagnetic resonance frequencies~\cite{Kittel1951,Keffer,Kamra2018}. For equivalent sublattices and assuming $\alpha\ll 1$ and $K\approx H_{0}M_{0}\ll J$, these read $\omega_{\textrm{p}\pm}\approx\left(1\pm\textrm{i}\alpha\sqrt{J/\left(4K\right)}\right)\left(\gamma H_{0}\pm\gamma/M\sqrt{4KJ}\right)$. Compared to the FM case, two resonance frequencies are found, and they are exchange enhanced by about a factor of $\sqrt{J/K}$. However, the lifetime of the excitations is reduced since the effective damping is also higher by a factor of $\sqrt{J/\left(4K\right)}$.
\begin{figure*}[htb]
\centering
 \includegraphics[width=2\columnwidth]{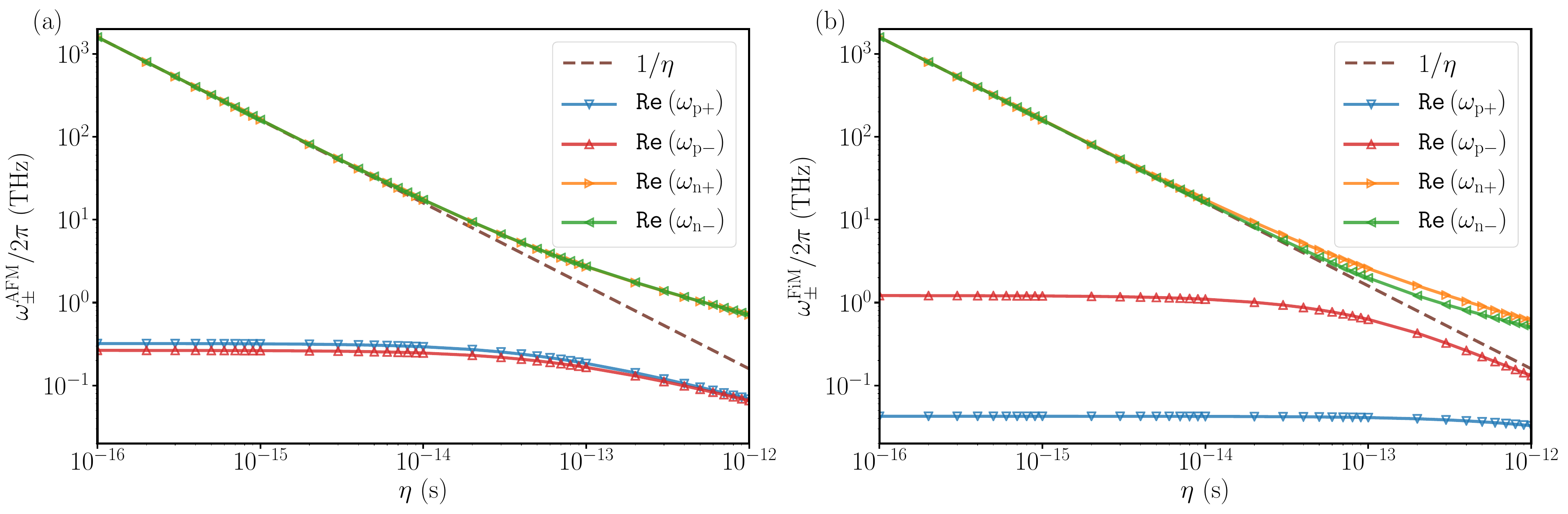}
\caption{(Color Online) 
Real part of the precession resonance frequencies as a function of inertial relaxation time $\eta$,  (a) for AFMs with $M_{A0} = M_{B0}=2\mu_{\textrm{B}}$ and (b) for FiMs with $M_{A0} = 5 M_{B0}=10\mu_{\textrm{B}}$. {  The other parameters are $ \gamma_A = \gamma_B  = 1.76 \times 10^{11}$~T$^{-1}$s$^{-1}$, $\alpha_{A} = \alpha_{B} = 0.05$, $K_A = K_B = 10^{-23}$~J, $J = 10^{-21}$~J, $ H_0 = 1$~T.}}  
\label{resonant_frequency}
\end{figure*}

In the presence of the inertial term, the resonance frequencies are found as a solution of a fourth-order equation. { The real and imaginary parts of the calculated frequencies are denoted by ${\tt Re}\left(\omega_{\rm p,n \pm}\right)$ and ${\tt Im}\left(\omega_{\rm p,n\pm}\right)$ for precession and nutation resonances, respectively.}
These have been calculated for an AFM and a FiM as a function of the relaxation time $\eta_{A}=\eta_{B}=\eta$ in Fig.~\ref{resonant_frequency}.  
In the absence of external field and damping, Eq.~\eqref{generic-equation} simplifies to a second-order equation in $\omega^{2}$. The precession resonance frequencies are given by $\omega_{\textrm{p}\pm}\approx\pm\gamma/M\sqrt{4KJ}\left(1+2{ \beta_{\rm AFM}}\right)^{-\frac{1}{2}}$ for $K\ll J$. It is important to note here that the relative strength of the inertial corrections is defined by the dimensionless parameter ${ \beta_{\rm AFM} = } { (}\eta\gamma/M_0{ )}J$, which is enhanced by a factor of $J/K$ as compared to $\beta_{\rm FM}$. 
The characteristic time scale of the exchange interactions typically falls into the fs range in AFMs which are ordered at room temperature ($\gamma J/M\approx 10^{13}$~s$^{-1}$ with the parameters used here), which is similar to the typical values of the inverse inertial relaxation time~\cite{Bhattacharjee2012,Li2015,neeraj2019experimental}. This explains the considerable decrease of the AFMR precession frequencies in Fig.~\ref{dissipation_antiferro}, while Fig.~\ref{resonant_frequency}(a) demonstrates that deviations from the non-inertial case already become observable for $\eta\approx1$~fs. This more pronounced inertial effect should also be observable if the resonance is measured by sweeping the external field, as in Ref.~\cite{Li2015}. The strongly asymmetric ($M_{A0} = 5 M_{B0}$) FiM in Fig.~\ref{resonant_frequency}(b) is characterized by a high-frequency { exchange} mode, strongly influenced by inertial effects as in the AFM, and a low-frequency mode which is less affected like in the FM.

The nutation resonance frequencies in the AFM can be expressed as $\omega_{\textrm{n}\pm}\approx\pm\sqrt{1+2{ \beta_{\rm AFM}}}/\eta$. Just as for the precession resonance, the correction factor arising due to the interplay between inertia and magnetic interactions is given by ${ \beta_{\rm AFM}}$, which is exchange enhanced compared to the FM case. This gives rise to an increase of the nutation frequencies, as demonstrated in Fig.~\ref{dissipation_antiferro}. For the FiM in Fig.~\ref{resonant_frequency}(b), the nutation frequency ${ {\tt Re}\left(\omega_{\textrm{n}+}\right)}$ belonging to the { exchange mode} ${ {\tt Re}\left(\omega_{\textrm{p}-}\right)}$ starts deviating from the low-inertia $\eta^{-1}$ asymptote at considerably lower frequencies than the FM-like nutation ${ {\tt Re}\left(\omega_{\textrm{n}-}\right)}$.

\begin{figure*}[htb]
\centering
 \includegraphics[width= 2\columnwidth]{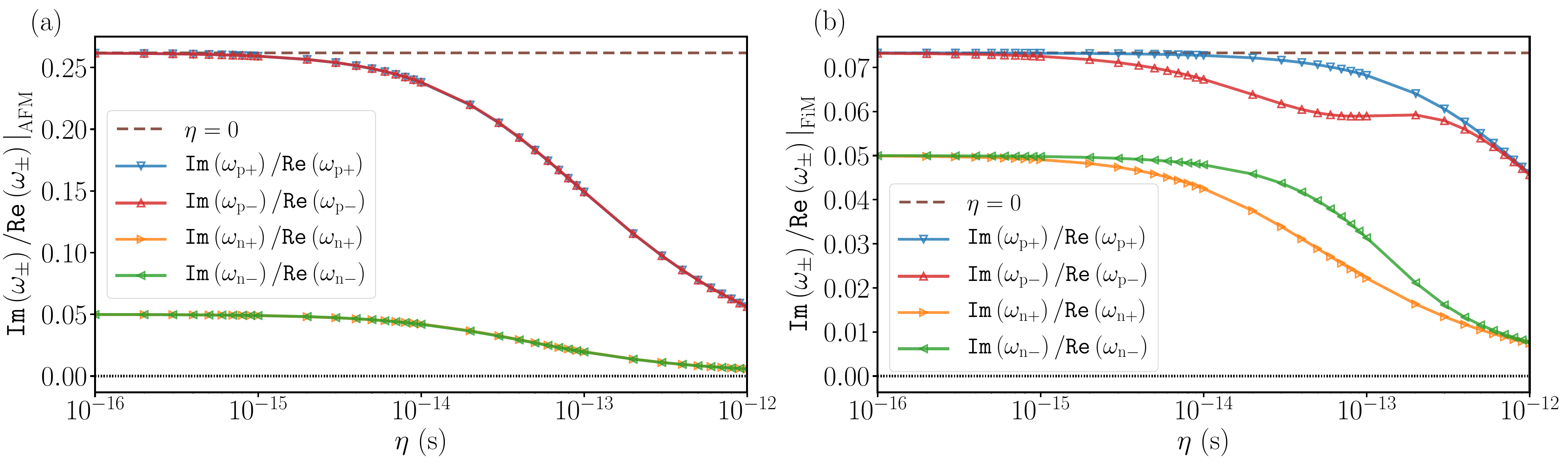}
\caption{(Color Online) Effective damping parameters of the resonance modes as a function of inertial relaxation time $\eta$, for (a) AFMs with $M_{A0} = M_{B0}=2\mu_{\textrm{B}}$ and (b) FiMs with $M_{A0} = 5 M_{B0}=10\mu_{\textrm{B}}$. 
{   The other parameters are $ \gamma_A = \gamma_B  = 1.76 \times 10^{11}$~T$^{-1}$s$^{-1}$, $\alpha_{A} = \alpha_{B} = 0.05$, $K_A = K_B = 10^{-23}$~J, $J = 10^{-21}$~J, $ H_0 = 1$~T.}} 
\label{effective_damping}
\end{figure*}

The effective damping parameters of the excitation modes, defined as the ratio of the imaginary to the real part of the frequencies, are shown in Fig.~\ref{effective_damping}. 
They no longer { coincide between precession and nutation} as in the FM case, since the exchange enhancement discussed above does not affect the nutation resonance. A reduction of the effective damping is observed with increasing inertial relaxation times, which becomes noticeable for ${ \beta_{\rm AFM}}=O\left(10^{-2}\right)$, just as in the case of the resonance frequencies. The considerable reduction of the effective damping compared to the Gilbert damping leads to sharper nutation resonance peaks as demonstrated in Fig.~\ref{dissipation_antiferro}, with higher intensities than for the FM. In the FiM, the { exchange} modes $\omega_{\textrm{n}+}$ and $\omega_{\textrm{p}-}$ start to become influenced at lower inertial relaxation times than the { FM} modes $\omega_{\textrm{n}-}$ and $\omega_{\textrm{p}+}$ ~\cite{Frank2012}. The difference between the effective damping parameters vanishes between { exchange} and { FM} modes for higher $\eta$, but it remains to be observable between precession and nutation modes.

{ \section{Conclusions}}

To conclude, we applied the ILLG equation to FMs and to two-sublattice AFMs and FiMs, and investigated the resonance frequencies using linear-response theory and computer simulations.
The precession frequencies are found to decrease with increasing inertial relaxation time and additional high-frequency nutation peaks become observable. Furthermore, the calculation of the resonance linewidth shows that the effect of inertia reduces the effective damping parameter. While in FMs these corrections scale with ${ \beta_{\rm FM} }=\eta\Omega_{0}$, in AFMs the dimensionless coupling between precession and nutation is given by ${ \beta_{\rm AFM}}={ (}\eta\gamma/M_0{ )}J$, which is typically several orders of magnitude higher. { Therefore, an antiferromagnetic system with higher exchange to anisotropy energy ratio and higher $\eta$ will be suitable to observe inertial effects. Such antiferromagnetic systems include NiO~\cite{Hutchings1972} and CrPt ~\cite{Besnus1973,Zhang2012JAP}, even though the characteristic inertial relaxation time $\eta$ is unknown. } The FiM is observed to interpolate between the FM and AFM limits. The reduced effective damping gives rise to particularly sharp and high-intensity nutation resonance peaks in AFMs, with frequencies comparable to the values already observed in FMs~\cite{Li2015,neeraj2019experimental}. These findings are expected to motivate the search for the signs of intrinsically inertial spin dynamics on ultrafast timescales using AFMR techniques.

{ \section*{Acknowledgments}}
We acknowledge financial support from the Alexander von Humboldt-Stiftung, the Deutsche Forschungsgemeinschaft via Project No. NO 290/5-1, and the National Research, Development, and Innovation Office of Hungary via Project No. K131938.

\appendix 
{ 
\section{Atomistic simulations of the ILLG equation}
\label{appendixA}
The inertial Landau-Lifshitz-Gilbert (ILLG) equation of motion, given in Eq.~(1) in the main text, can be rewritten for the normalized spin $\bm{s_i}(t)=\bm{M}_{i}(t)/M_{i0}$ as \cite{Mondal2017Nutation}
\begin{align}
    \partial_{t} \bm{s}_i = - \gamma_i \bm{s}_i\times \bm{H}_i + \alpha_i \bm{s}_i \times \partial_{t} \bm{s}_i + \eta_i \bm{s}_i\times \partial_{tt} \bm{s}_i\,.
    \label{LLGN_equation}
\end{align}
The first term denotes precession of the spins around an effective field $\bm{H}_i$, the second term corresponds to a transverse relaxation of the spins, and the last term defines the inertial dynamics \cite{Ciornei2011}.
The ILLG equation can be rewritten from the implicit form of Eq.~(\ref{LLGN_equation}) to an explicit differential equation which can be solved numerically without iterations. By taking a scalar product of Eq.~(\ref{LLGN_equation}) with $\bm{s}_i$ it is easy to see that the length of the spin remains conserved in the ILLG equation, i.e.,  $\partial_{t}\vert \bm{s}_i\vert^2 = 0$ and $\bm{s}_i\cdot\partial_{t} \bm{s}_i = 0$. Furthermore, we use
\begin{align}
\label{cond.1}
    \bm{s}_i\times \left(\bm{s}_i\times \partial_{tt} \bm{s}_i\right) &= \bm{s}_i\left( \bm{s}_i\cdot\partial_{tt} \bm{s}_i\right) - \partial_{tt} \bm{s}_i\,, \\
    \partial_{t}\underbrace{\left(\bm{s}_i\cdot\partial_{t} \bm{s}_i\right)}_{=0} &= \left(\partial_{t} \bm{s}_i\right)^2 + \bm{s}_i\cdot\partial_{tt} \bm{s}_i\,.
    \label{cond.2}
\end{align}
By multiplying Eq.~(\ref{LLGN_equation}) by $\bm{s}_i\times$ and using the conditions Eqs.~(\ref{cond.1}) and (\ref{cond.2}), we obtain the explicit equation of motion (cf. Ref.~\cite{Kikuchi})
\begin{align}
    \partial_{tt} \bm{s}_i & = -\frac{\gamma_i}{\eta_i} \bm{s}_i\times\left(\bm{s}_i \times \bm{H}_i\right) - \frac{\alpha_i}{\eta_i} \partial_{t} \bm{s}_i - \frac{1}{\eta_i} \bm{s}_i \times \partial_{t} \bm{s}_i \nonumber\\
    & - \bm{s}_i\left(\partial_{t} \bm{s}_i\right)^2 = \bm{F}_i\left(\bm{s},\partial_{t} \bm{s}, t\right) \,.\label{eILLG}
\end{align}
Note that a second-order explicit differential equation is obtained because of the inertial term, while the LLG equation is of first order. 
With the definition $\bm{p}_i = \partial_{t} \bm{s}_i$, we can convert the second-order differential equation into a system of first-order differential equations as follows:
\begin{align}
   \partial_{tt} \bm{s}_i =\partial_{t} \bm{p}_i &= \bm{F}_i\left( \bm{s},\bm{p},t\right)\,,\\
   \partial_{t} \bm{s}_i  = \bm{p}_i &= \bm{G}_i\left(\bm{s},\bm{p},t\right)\,.
\end{align}
It is obvious that one has to solve six coupled differential equations of first order per lattice site $i$. We numerically solve these equations with Heun's method~\cite{nowak2007handbook}, where the predictor steps are
\begin{align}
\bar{\bm{s}}_i &=\bm{s}_i(t) + \Delta t\bm{G}_i\left(\bm{s},\bm{p},t\right)\,, \\
 \bar{\bm{p}}_i &=\bm{p}_i(t) + \Delta t\bm{F}_i\left(\bm{s},\bm{p},t\right)\,,
\end{align}
and the corrector steps are implemented as 
\begin{align}
    \bm{s}_i(t+\Delta t) &= \bm{s}_i(t) + \frac{\Delta t}{2}\left[\bm{G}_i\left(\bm{s},\bm{p},t\right) + \bm{G}_i\left(\bar{\bm{s}} , \bar{\bm{p}}, t+\Delta t \right) \right]\,, \\
    \bm{p}_i(t+\Delta t) &= \bm{p}_i(t) + \frac{\Delta t}{2}\left[\bm{F}_i\left(\bm{s},\bm{p},t\right) + \bm{F}_i\left(\bar{\bm{s}} , \bar{\bm{p}}, t+\Delta t \right) \right]\,.
\end{align}
In order to calculate the resonance curves, we employed a circularly polarized field $\bm{h}(t) \sim e^{\textrm{i}\omega t}$ in the $xy$ plane in addition to the static magnetic field $ \bm{H}_0$ along the $z$ direction, and solved the equations of motion for one and two spins by starting from the equilibrium state along the $z$ direction.

By multiplying Eq.~\eqref{eILLG} by $M_{i0}\eta_{i}\partial_{t} \bm{s}_i/\gamma_{i}$, summing over the sublattices, and rearranging the terms, one arrives at
\begin{align}
& \partial_{t}\left(\sum_{i}\frac{M_{i0}\eta_{i}}{2\gamma_{i}}\left(\partial_{t}\bm{s}_i\right)^{2}+\mathcal{F}\right)=\sum_{i}\partial_{t}M_{i0}\partial_{t}\bm{s}_i\bm{h}_i \nonumber\\
& -\sum_{i}\alpha_{i}\frac{M_{i0}}{\gamma_{i}}\left(\partial_{t}\bm{s}_i\right)^{2}\,.\label{energybalance}
\end{align}
The left-hand side of Eq.~\eqref{energybalance} describes the change of rate of the energy of the system, consisting of a kinetic part and a potential part $\mathcal{F}$. The former sheds light on the meaning of $\eta_{i}$ as an inertial parameter. The right-hand side consists of the power loss due to damping processes, which is compensated by the external driving force in a steady state. Accordingly, we computed the dissipated power using $P = \sum_i M_{i0}\partial_{t}\bm{s}_i\cdot\bm{h}_i$.

\section{Calculation of the linear response in ferromagnets}
\label{appendixB}
In ferromagnets, we consider that the initial  magnetization 
points towards the $z$ direction, such that the magnetization is expanded as $\bm{M} = M_{0} \hat{\bm{e}}_z + \bm{m}({t})$ in linear order. 
The considered dynamical field is denoted by $\bm{h}(t)$.
Using the effective field in the main text, the linearized ILLG equation can be written
in the following way:
\begin{widetext}
\begin{align}
     \partial_{t} \bm{m}  & =
    - \gamma \left[\underbrace{M_{\rm 0}\hat{\bm{e}}_z \times H_{\rm 0}\hat{\bm{e}}_z}_{\rm = \,\, 0} + \underbrace{M_{\rm 0}\hat{\bm{e}}_z \times \frac{2K}{M_0}\hat{\bm{e}}_z}_{\rm =\,\, 0} + M_{\rm 0}\hat{\bm{e}}_z \times \bm{h}(t) + \bm{m}(t)\times H_{\rm 0}\hat{\bm{e}}_z + \bm{m}(t)\times \frac{2K}{M_0}\hat{\bm{e}}_z+ \underbrace{\bm{m}(t) \times \bm{h}(t)}_{\rm negligible} \right] \nonumber\\
    & + \frac{\alpha}{M_{\rm 0}}\left[M_{\rm 0}\hat{\bm{e}}_z \times  \frac{\partial \bm{m}}{\partial t} + \underbrace{\bm{m}\times \frac{\partial \bm{m}}{\partial t}}_{\rm negligible}\right] + \frac{\eta}{M_{\rm 0}}\left[M_{\rm 0}\hat{\bm{e}}_z \times  \frac{\partial^2 \bm{m}}{\partial t^2} + \underbrace{\bm{m}\times \frac{\partial^2 \bm{m}}{\partial t^2}}_{\rm negligible}\right]\,.
\end{align}
\end{widetext}
Thus, we obtain the following two equations for the transversal components:
\begin{align}
    \partial _t m_x & =\gamma M_0 h_y - \gamma H_0 m_y - \frac{2\gamma K}{M_0}m_y  - \alpha \partial_t m_y   - \eta \partial_{tt} m_y\,, \label{eqFM1} \\
    \partial _t m_y & = - \gamma M_0 h_x + \gamma H_0 m_x + \frac{2\gamma K}{M_0}m_x+ \alpha \partial_t m_x + \eta \partial_{tt} m_x\,. \label{eqFM2}
\end{align}
We define $\Omega_{0} = \gamma/M_0 \left( H_0 M_0 + 2K\right) $ as in the main text.
Therefore, Eqs.~\eqref{eqFM1} and \eqref{eqFM2} can be recast as
\begin{align}
    h_x & = \frac{1}{\gamma M_0} \left[\Omega_{0}m_x +\alpha \partial_t m_x + \eta \partial_{tt} m_x - \partial _t m_y  \right]\,,\\
    h_y & = \frac{1}{\gamma M_0}\left[\Omega_{0} m_y+\alpha \partial_t m_y + \eta \partial_{tt} m_y + \partial _t m_x  \right]\,.
\end{align}
In matrix form we write
\begin{align}
    \begin{pmatrix}
h_x\\
h_y
\end{pmatrix} & = \frac{1}{\gamma M_0} \begin{pmatrix}
\Omega_{0} +\alpha \partial_t + \eta \partial_{tt}   & - \partial _t \\
\partial _t & \Omega_{0} +\alpha \partial_t + \eta \partial_{tt} 
\end{pmatrix}\begin{pmatrix}
m_x\\
m_y
\end{pmatrix}\,.
\end{align}
We switch to the circularly polarized basis, $m_{\pm} = m_x \pm \textrm{i}m_y$ and $h_{\pm} = h_x \pm \textrm{i}h_y$, where the equations decouple,
\begin{align}
&  \gamma M_0  \begin{pmatrix}
h_+\\
h_-
\end{pmatrix}  = \nonumber\\
& \begin{pmatrix}
\Omega_{0} +\alpha \partial_t + \eta \partial_{tt} + \textrm{i}\partial _t  & 0 \\
0 & \Omega_{0} +\alpha \partial_t + \eta \partial_{tt} - \textrm{i}\partial _t
\end{pmatrix}\begin{pmatrix}
m_+\\
m_-
\end{pmatrix}\,.
\end{align}
For the time dependence we consider $h_\pm = h e^{\pm \textrm{i}\omega t}$, describing two types of polarization with opposite handedness. We assume $m_\pm = m e^{\pm \textrm{i}\omega t}$. Thus, we have
\begin{align}
    h e^{\textrm{i}\omega t} & = \frac{1}{\gamma M_0} \left(\Omega_{0} + \textrm{i} \alpha \omega  - \eta \omega^2 - \omega  \right) m e^{\textrm{i}\omega t} \nonumber\\
       \Rightarrow m_+ & = \frac{\gamma M_0}{\Omega_{0} + i \alpha \omega  - \eta \omega^2 - \omega }h e^{i\omega t}\,, \label{chiFM+}\\
    h e^{-\textrm{i}\omega t} & = \frac{1}{\gamma M_0}\left(\Omega_{0} - \textrm{i} \alpha \omega  - \eta \omega^2 - \omega  \right) m e^{-\textrm{i}\omega t} \nonumber\\
     \Rightarrow m_- & = \frac{\gamma M_0}{\Omega_{0} - \textrm{i} \alpha \omega  - \eta \omega^2 - \omega }h e^{-\textrm{i}\omega t}\,.
\end{align}
This leads to the susceptibility given in Eq.~\eqref{suscFM}.
Its real and imaginary parts are derived as
\begin{align}
    {\tt Re}(\chi_{\pm}) & = \gamma M_0 \frac{\Omega_{0}  - \omega - \eta \omega^2 }{(\Omega_{0} - \omega - \eta \omega^2)^2 + \alpha^2\omega^2}\,, \label{ResuscFM} \\ 
    {\tt Im}(\chi_{\pm}) & = \pm\gamma M_0 \frac{\alpha \omega}{(\Omega_{0} - \omega - \eta \omega^2)^2 + \alpha^2\omega^2}\,. \label{ImsuscFM}
\end{align}

The dissipated power can be calculated according to its definition based on Eq.~\eqref{energybalance},
\begin{align}
    P  & = \partial_{t}\bm{m}\cdot \bm{h} \nonumber\\
    & = \left(\partial_{t}m_xh_x + \partial_{t}m_yh_y\right)\nonumber\\
     & = \frac{1}{2}\left( \partial_{t}m_+h_- + \partial_{t}m_-h_+ \right) \nonumber\\
       &  = \frac{\textrm{i}\omega}{2} \left(\chi_+ - \chi_-\right)\vert h \vert^2\nonumber\\
          & = \frac{\textrm{i}\omega}{2} \left(\frac{-2\textrm{i} \alpha \omega \gamma M_0}{(\Omega_{0} - \omega - \eta \omega^2)^2 + \alpha^2\omega^2 }\right)\vert h \vert^2 \nonumber\\
          & = \omega {\tt Im}(\chi_+)\vert h \vert^2 \,.
\end{align}

The positions and the linewidths of the resonance peaks may be analyzed  by finding the poles of the susceptibility in Eq.~\eqref{chiFM+},
\begin{align}
\omega & =\frac{1}{2\eta}\left[-\left(1-\textrm{i}\alpha\right)\pm\sqrt{\left(1-\textrm{i}\alpha\right)^2+4\beta_{\textrm{FM}}}\right]\nonumber\\
& =\frac{1}{2\eta}\left[-1\pm a+\textrm{i}\alpha\left(1\mp a^{-1}\right)\right]\,, \label{FMfreq}
\end{align}
where $\beta_{\textrm{FM}}=\eta\Omega_{0}$ and $a$ is the single positive real solution of the fourth-order equation
\begin{align}
a^{4}-\left(1-\alpha^{2}+4\beta_{\textrm{FM}}\right)a^{2}-\alpha^{2}=0\,.
\end{align}

For $\beta_{\textrm{FM}}\ll 1$, one has $a=1+2\beta_{\textrm{FM}}+O\left(\beta_{\textrm{FM}}^{2}\right)$. For the real parts of the frequencies, corresponding to the peak positions, one obtains $\omega_{\textrm{p}}\approx \Omega_{0}\left(1-\beta_{\textrm{FM}}\right)$ and $\omega_{\textrm{n}}\approx -1/\eta-\Omega_{0}\left(1-\beta_{\textrm{FM}}\right)$, as described in the main text. Note that the latter expression agrees with Eq.~(14) in Ref.~\cite{cherkasskii2020nutation}, but the correction terms are different from Ref.~\cite{Olive2015}, where $\omega_{\textrm{n}}=-\sqrt{1+\beta_{\rm FM}}/\eta\approx -1/\eta-\Omega_{0}/2\left(1-\beta_{\textrm{FM}/4}\right)$ was suggested. It is apparent from Eq.~\eqref{FMfreq} that effective damping parameter, i.e. the ratio of the imaginary and the real parts of the frequency, is $\alpha a^{-1} \approx\alpha\left(1-2\beta_{\textrm{FM}}\right)$ both for the precession and the nutation peaks. The full width of the resonance peaks at half maximum can be expressed as $\Delta \omega=\omega_{1}-\omega_{2}$, which frequencies satisfy
\begin{widetext}
\begin{align}
\Omega_{0} - \omega_{1} - \eta \omega_{1}^2 & = -\alpha\omega_{1}\,, \\
\Omega_{0} - \omega_{2} - \eta \omega_{2}^2 & = \alpha\omega_{2}\,. 
\end{align}
The ratio of the linewidth and the peak position is given by
\begin{align}
    \frac{\Delta\omega}{\omega_{\textrm{p}}} & = \frac{\left(\Omega_0+\alpha\Omega_0\left(1-\beta_{\textrm{FM}}\right)\right) - \eta\left(\Omega_0+\alpha\Omega_0\right)^2 -\left(\Omega_0-\alpha\Omega_0\left(1-\beta_{\textrm{FM}}\right)\right) + \eta\left(\Omega_0-\alpha\Omega_0\right)^2 }{\Omega_0 \left(1 -\beta_{\textrm{FM}}\right)} \nonumber\\
    & = \frac{2\alpha \Omega_0 - 6\alpha\beta_{\textrm{FM}} \Omega_0 }{\Omega_0 \left(1 -\beta_{\textrm{FM}}\right)}
     = 2\alpha \frac{1 - 3 \beta_{\textrm{FM}}}{1 -  \beta_{\textrm{FM}}}
      \approx 2\alpha \left(1 - 2 \beta_{\textrm{FM}}\right) \,
\end{align}
\end{widetext}
for the precession resonance, confirming that dividing the half-width at half maximum by the resonance frequency is approximately equal to the effective damping parameter described above. 

\section{Calculation of the linear response in two-sublattice antiferromagnets and ferrimagnets}
\label{appendixC} 
We expand the magnetization around the equilibrium direction in small deviations, $\bm{M}_{A} = M_{A0}\hat{\bm{e}}_z+\bm{m}_{A}$ and $\bm{M}_{B} = -M_{B0}\hat{\bm{e}}_z+\bm{m}_{B}$, which are induced by the transverse external field $\bm{h}_{A/B}(t)$. The linearized ILLG equation for the two sublattices reads
\begin{widetext}
\begin{align}
     \partial_t \bm{m}_{A} & = -\frac{\gamma_A}{M_{A0}}\left[-( H_0 M_{A0} + 2K_A+ J)m_{Ax}\hat{\bm{e}}_y
     + ( H_0 M_{A0}+ 2K_A+ J)m_{Ay}\hat{\bm{e}}_x  
          \right] + \frac{ \gamma_A}{M_{B0}} \left[
    Jm_{Bx}\hat{\bm{e}}_y - Jm_{By}\hat{\bm{e}}_x\right] \nonumber\\
    & - \gamma_A M_{A0} \left(h_{Ax}\hat{\bm{e}}_y- h_{Ay}\hat{\bm{e}}_x\right) + \alpha_{A} \left(\partial_t m_{Ax}\hat{\bm{e}}_y - \partial_t m_{Ay}\hat{\bm{e}}_x\right)   + \eta_{A}\left(\partial_{tt} m_{Ax}\hat{\bm{e}}_y - \partial_{tt} m_{Ay}\hat{\bm{e}}_x \right)\,, \\
     \partial_t \bm{m}_{B} & = -\frac{\gamma_B}{M_{B0}}\left[-( H_0 M_{B0} - 2K_B- J)m_{Bx}\hat{\bm{e}}_y
     + ( H_0 M_{B0}- 2K_B- J)m_{By}\hat{\bm{e}}_x  
     \right] - \frac{ \gamma_B}{M_{A0}} \left[
    Jm_{Ax}\hat{\bm{e}}_y - Jm_{Ay}\hat{\bm{e}}_x\right] \nonumber\\
    & + \gamma_B M_{B0} \left(h_{Bx}\hat{\bm{e}}_y- h_{By}\hat{\bm{e}}_x\right) - \alpha_{B} \left(\partial_t m_{Bx}\hat{\bm{e}}_y - \partial_t m_{By}\hat{\bm{e}}_x\right)   - \eta_{B}\left(\partial_{tt} m_{Bx}\hat{\bm{e}}_y - \partial_{tt} m_{By}\hat{\bm{e}}_x \right)\,.
\end{align}
For the $x$ and $y$ components we obtain
\begin{align}
  \gamma_A  M_{A0}h_{Ay} & = \frac{ \gamma_A }{M_{A0}}
( H_0 M_{A0}+ 2K_A+ J)m_{Ay}  + \frac{ \gamma_A}{M_{B0}}   Jm_{By} + \alpha_{A}   \partial_t m_{Ay}   + \eta_{A} \partial_{tt} m_{Ay} + \partial_t m_{Ax}\,,\\
 \gamma_A  M_{A0} h_{Ax} & = \frac{ \gamma_A }{M_{A0}}
( H_0 M_{A0}+ 2K_A+ J)m_{Ax}  + \frac{ \gamma_A}{M_{B0}}   Jm_{Bx} + \alpha_{A}   \partial_t m_{Ax}   + \eta_{A} \partial_{tt} m_{Ax} - \partial_t m_{Ay}\,,\\
  \gamma_B  M_{B0}h_{By} & = \frac{ \gamma_B }{M_{B0}}
(- H_0 M_{B0}+ 2K_B+ J)m_{By}  + \frac{ \gamma_B}{M_{A0}}   Jm_{Ay} + \alpha_{B}   \partial_t m_{By}   + \eta_{B} \partial_{tt} m_{By} - \partial_t m_{Bx}\,,\\
 \gamma_B  M_{B0} h_{Bx} & = \frac{ \gamma_B }{M_{B0}}
(- H_0 M_{B0}+ 2K_B+ J)m_{Bx}  + \frac{ \gamma_B}{M_{A0}}   Jm_{Ax} + \alpha_{B}   \partial_t m_{Bx}   + \eta_{B} \partial_{tt} m_{Bx} + \partial_t m_{By}\,.
\end{align}
In the circularly polarized basis with $m_{A/B\pm} = m_{A/Bx} \pm \textrm{i} m_{A/By},h_{A/B\pm} = h_{A/Bx} \pm \textrm{i} h_{A/By}$ and defining $\Omega_A = \gamma_A /M_{A0}( H_0 M_{A0}+ 2K_A+ J), \Omega_B = \gamma_B /M_{B0}
( J + 2K_B -  H_0 M_{B0})$, we obtain
\begin{align}
     \gamma_A  M_{A0} h_{A\pm} & = \left(\Omega_A +\alpha_{A}\partial_t +\eta_{A}\partial_{tt} \pm \textrm{i} \partial_t  \right)m_{A\pm} + \frac{ \gamma_A }{M_{B0}} J m_{B\pm}\,, \\
     \gamma_B  M_{B0} h_{B\pm} & = \left(\Omega_B +\alpha_{B}\partial_t +\eta_{B}\partial_{tt} \mp \textrm{i} \partial_t  \right)m_{B\pm} + \frac{ \gamma_B }{M_{A0}} J m_{A\pm}\,.
\end{align}
The four equations of motion are separated into two pairs of coupled equations for the $+$ and $-$ components. In matrix formalism we have
\begin{align}
    \begin{pmatrix}
      h_{A\pm}\\
       h_{B\pm}
    \end{pmatrix} & = \begin{pmatrix}
      \dfrac{1}{ \gamma_A M_{A0}}\left(\Omega_A +\alpha_{A}\partial_t +\eta_{A}\partial_{tt} \pm \textrm{i} \partial_t  \right) & \dfrac{1}{M_{A0}M_{B0}} J\\
     \dfrac{1}{M_{A0}M_{B0}}J &  \dfrac{1}{ \gamma_B M_{B0}}\left(\Omega_B +\alpha_{B}\partial_t +\eta_{B}\partial_{tt} \mp \textrm{i} \partial_t  \right)
    \end{pmatrix}\begin{pmatrix}
      m_{A\pm}\\
      m_{B\pm}
    \end{pmatrix}\,.
\end{align}
By substituting the time dependence $h_{A/B\pm},m_{A/B\pm}\propto\textrm{e}^{\pm\textrm{i}\omega t}$ we have
\begin{align}
    \begin{pmatrix}
     h_{A\pm}\\
     h_{B\pm}
    \end{pmatrix} & = \begin{pmatrix}
      \dfrac{1}{ \gamma_A M_{A0}}\left(\Omega_A \pm \textrm{i}\omega\alpha_{A} -\eta_{A}\omega^2-\omega  \right) &  \dfrac{1}{M_{A0}M_{B0}}J\\
      \dfrac{1}{M_{A0}M_{B0}}J &   \dfrac{1}{ \gamma_B M_{B0}}\left(\Omega_B \pm \textrm{i}\omega\alpha_{B} -\eta_{B}\omega^2 +\omega  \right)
    \end{pmatrix}\begin{pmatrix}
      m_{A\pm}\\
      m_{B\pm}
    \end{pmatrix}\,.
\end{align}
We introduce the definition $\Delta_\pm = \left(\gamma_A M_{A0} \gamma_B M_{B0}\right)^{-1}\left(\Omega_A \pm i\omega\alpha_{A} -\eta_{A}\omega^2-\omega  \right)\left(\Omega_B \pm i\omega\alpha_{B} -\eta_{B}\omega^2 +\omega  \right) - J^2/\left(M_{A0}^2 M_{B0}^2\right)$ for the determinant of the matrix above. The susceptibility tensor is obtained by matrix inversion,
\begin{align}
    \begin{pmatrix}
     m_{A\pm}\\
     m_{B\pm}
    \end{pmatrix} & = \frac{1}{\Delta_\pm}\begin{pmatrix}
     \dfrac{1}{ \gamma_B M_{B0}}\left(\Omega_B \pm i\omega\alpha_{B} -\eta_{B}\omega^2 +\omega  \right) &  -\dfrac{1}{M_{A0}M_{B0}}J\\
     - \dfrac{1}{M_{A0}M_{B0}}J &   \dfrac{1}{ \gamma_A M_{A0}}\left(\Omega_A \pm i\omega\alpha_{A} -\eta_{A}\omega^2-\omega  \right)
    \end{pmatrix}\begin{pmatrix}
      h_{A\pm}\\
      h_{B\pm}
    \end{pmatrix}
    & = \chi_{\pm}^{AB}\begin{pmatrix}
      h_{A\pm}\\
      h_{B\pm}
    \end{pmatrix}\,,
\end{align}
as also given in Eq.~\eqref{suscAFM}.

Similarly to the ferromagnet, we calculate the dissipated power from Eq.~\eqref{energybalance} as
\begin{align}
    P_{AB} & =  \partial_{t}\bm{m}_A\cdot \bm{h}_A + \partial_{t}\bm{m}_B\cdot \bm{h}_B \nonumber\\
    & = \frac{1}{2}\Big[ \partial_{t}m_{A+}h_{A-}   + \partial_{t}m_{A-}h_{A+} + \partial_{t}m_{B+}h_{B-}   + \partial_{t}m_{B-}h_{B+}\Big]\nonumber\\
    & = \frac{\textrm{i}\omega}{2}\Big[\frac{1}{\Delta_+} \left(\frac{1}{ \gamma_B M_{B0}}\left(\Omega_B + \textrm{i}\omega\alpha_{B} -\eta_{B}\omega^2 +\omega  \right)h_{A+} - \frac{1}{M_{A0}M_{B0}}Jh_{B+}\right)h_{A-}  \nonumber\\
    & -\frac{1}{\Delta_-} \left(\frac{1}{ \gamma_B M_{B0}}\left(\Omega_B - \textrm{i}\omega\alpha_{B} -\eta_{B}\omega^2 +\omega  \right)h_{A-} - \frac{1}{M_{A0}M_{B0}}Jh_{B-}\right)h_{A+} \nonumber\\
    &  + \frac{1}{\Delta_+}\left( - \frac{1}{M_{A0}M_{B0}}Jh_{A+} + \frac{1}{ \gamma_A M_{A0}}\left(\Omega_A + \textrm{i}\omega\alpha_{A} -\eta_{A}\omega^2 -\omega  \right)h_{B+}\right)h_{B-} \nonumber\\
    & - \frac{1}{\Delta_-}\left( - \frac{1}{M_{A0}M_{B0}}Jh_{A-} + \frac{1}{ \gamma_A M_{A0}}\left(\Omega_A - \textrm{i}\omega\alpha_{A} -\eta_{A}\omega^2 -\omega  \right)h_{B-}\right)h_{B+}\Big]\nonumber\\
    & = \frac{\omega^2 \vert h_A\vert^2}{\gamma_B M_{B0}}\left[    \frac{ \left( \gamma_A M_{A0}\gamma_B M_{B0}\right)^{-1} \alpha_{A}\left[\left(\Omega_B-\eta_{B}\omega^2+\omega\right)^2+\omega^2\alpha_{B}^2\right]+J^2/\left(M_{A0}^2M_{B0}^2\right)\alpha_{B}}{\Delta_+\Delta_-}\right]    \nonumber\\
    & + \frac{\omega^2 \vert h_B\vert^2}{\gamma_A M_{A0}}\left[    \frac{ \left( \gamma_A M_{A0}\gamma_B M_{B0}\right)^{-1} \alpha_{B}\left[\left(\Omega_A-\eta_{A}\omega^2-\omega\right)^2+\omega^2\alpha_{A}^2\right]+J^2/\left(M_{A0}^2M_{B0}\right)^2\alpha_{A}}{\Delta_+\Delta_-}\right]\nonumber\\
    & - \frac{2\omega^2 J \vert h_A h_B \vert}{ \gamma_A M_{A0}^2 \gamma_B M_{B0}^2}\left[\frac{\left(\Omega_A\alpha_{B}+\Omega_B\alpha_{A}\right) + \left(\alpha_{A}-\alpha_{B}\right)\omega-\left(\eta_{A}\alpha_{B}+\eta_{B}\alpha_{A}\right)\omega^2}{\Delta_+\Delta_-}\right]\,.
\label{two-sub-dissipation}
\end{align}

As discussed in the main text, the peak positions and the linewidths may be understood by finding the nodes of the determinant $\Delta_{\pm}$,
\begin{align}
    & \left( \Omega_A \pm \textrm{i}\omega \alpha_{A} - \eta_{A}\omega^2-\omega \right)\left( \Omega_B \pm \textrm{i}\omega \alpha_{B} - \eta_{B}\omega^2 +\omega\right)  - \frac{ \gamma_A\gamma_B}{M_{A0} M_{B0}}J ^2  = 0\nonumber\\
    & \Rightarrow \underbrace{ \eta_{A}\eta_{B}
    }_{=a_{\pm}}\omega ^4+ \underbrace{\left[\mp \textrm{i} \left(\alpha_{A}\eta_{B}+\alpha_{B}\eta_{A}\right)- \left( \eta_{A} -\eta_{B} \right)\right]}_{=b_{\pm}} \omega^3 \nonumber\\
    & + \underbrace{\left[-1 \pm \textrm{i}(\alpha_{A} - \alpha_{B}) - (\Omega_A\eta_{B} + \Omega_{B} \eta_{A}) - \alpha_{A}\alpha_{B} \right]}_{=c_{\pm}}\omega^2 \nonumber\\
    & + \underbrace{\left[ \left(\Omega_A - \Omega_B\right) \pm \textrm{i}\left(\alpha_{B}\Omega_A + \alpha_{A}\Omega_B\right)\right]}_{=d_{\pm}}\omega  +\underbrace{\Omega_{A}\Omega_{B}-\frac{ \gamma_A\gamma_B}{M_{A0}M_{B0}}J^2}_{=e_{\pm}} = 0\,.\label{detAFM}
\end{align}
\end{widetext}

The fourth-order equation~\eqref{detAFM} may be solved in a closed form. However, in order to arrive at solutions which have a simpler form, we consider the antiferromagnet with identical sublattices, $M_{A0}=M_{B0}=M_{0}$, $\alpha_{A}=\alpha_{B}=\alpha$, $\eta_{A}=\eta_{B}=\eta$, and $K_{A}=K_{B}=K$. Furthermore, we assume $\alpha\ll 1$ and $M_{0}H_{0},K\ll J$, as is typical in most systems. Consequently, we will treat the terms proportional to the damping and the external field in first-order perturbation theory, leading to
\begin{align}
&\eta^{2}\omega^{4}-\left(1+2\eta\frac{\gamma}{M_{0}}\left(J+2K\right)\right)\omega^{2}-\textrm{i}2\alpha\eta\omega_{\left(0\right)}^{3}+2\gamma H_{0}\omega_{\left(0\right)} \nonumber\\
& + \textrm{i}2\alpha\frac{\gamma}{M_{0}}\left(J+2K\right)\omega_{\left(0\right)}+\frac{\gamma^{2}}{M^{2}_{0}}\left(J+2K\right)^{2}-\gamma^{2}\left(H_{0}\right)^{2} \nonumber\\
& -\frac{\gamma^{2}}{M^{2}_{0}}J^{2}=0\,,\label{detAFMsimple}
\end{align}
where $\omega_{\left(0\right)}$ is the solution for $\alpha=0$ and $H_{0}=0$, and we only treat $\Delta_{+}$ for simplicity since $\Delta_{-}$ may be obtained by complex conjugation. Equation~\eqref{detAFMsimple} is a second-order equation in $\omega^{2}$, the solutions of which are simple to express. Expanding them up to first order in $\alpha$ and $H_{0}$ for consistency with the order of the perturbation, and also in first order in $K/J\ll 1$, one obtains
\begin{widetext}
\begin{align}
\omega_{\textrm{p}\pm}\approx&\pm\frac{\gamma}{M_{0}}\frac{\sqrt{4K\left(J+K\right)}}{\sqrt{1+2\eta\dfrac{\gamma}{M_{0}}\left(J+2K\right)}}+\frac{1}{\sqrt{1+2\eta\dfrac{\gamma}{M_0}\left(J+2K\right)}}\nonumber\\
&\times\frac{\left|\omega_{(0)}\right|}{\dfrac{\gamma}{M_{0}}\sqrt{4K\left(J+K\right)}}\left[\gamma H_{0}+\textrm{i}\alpha\left(\frac{\gamma}{M_{0}}\left(J+2K\right)-\eta\omega_{(0)}^{2}\right)\right]\,, \\
\omega_{\textrm{n}\pm}\approx&\pm\frac{1}{\eta}\sqrt{1+2\eta\dfrac{\gamma}{M_{0}}\left(J+2K\right)}\left(1-\frac{\eta^{2}\dfrac{\gamma^{2}}{M^{2}_{0}}4K\left(J+K\right)}{2\left[1+2\eta\dfrac{\gamma}{M_{0}}\left(J+2K\right)\right]^{2}}\right)\nonumber\\
&-\frac{\eta\left|\omega_{(0)}\right|}{\left[1+2\eta\dfrac{\gamma}{M_{0}}\left(J+2K\right)\right]^{\frac{3}{2}}}\left[\gamma H_{0}+\textrm{i}\alpha\left(\frac{\gamma}{M_{0}}\left(J+2K\right)-\eta\omega_{(0)}^{2}\right)\right]\,, 
\end{align}
for the precession and the nutation frequencies, respectively. Substituting in $\left|\omega_{(0)}\right|$ from the leading term in the expression into the perturbative terms, one arrives at
\begin{align}
\omega_{\textrm{p}\pm}\approx&\pm\frac{\gamma}{M_{0}}\frac{\sqrt{4K\left(J+K\right)}}{\sqrt{1+2\eta\dfrac{\gamma}{M_{0}}\left(J+2K\right)}}\nonumber\\
&+\frac{1}{1+2\eta\dfrac{\gamma}{M_{0}}\left(J+2K\right)}\left[\gamma H_{0}+\textrm{i}\alpha\frac{\gamma}{M_{0}}\left(J+2K\right)\right]\,, \label{AFMprec}\\
\omega_{\textrm{n}\pm}\approx&\pm\frac{1}{\eta}\sqrt{1+2\eta\dfrac{\gamma}{M_{0}}\left(J+2K\right)}\left(1-\frac{\eta^{2}\dfrac{\gamma^{2}}{M^{2}_{0}}4K\left(J+K\right)}{2\left[1+2\eta\dfrac{\gamma}{M_{0}}\left(J+2K\right)\right]^{2}}\right)\nonumber\\
& -\frac{1}{1+2\eta\dfrac{\gamma}{M_{0}}\left(J+2K\right)}\left[\gamma H_{0}-\textrm{i}\alpha\left(\frac{1}{\eta}+\frac{\gamma}{M_{0}}\left(J+2K\right)\right)\right]\,. \label{AFMnut}
\end{align}
\end{widetext}

The leading-order terms for $H_{0},\alpha=0$ and using $J+K\approx J$ are also reported in the main text. As discussed there, in the antiferromagnet the corrections caused by the inertial dynamics surpass in magnitude those in the ferromagnet, since the characteristic dimensionless parameter $\beta_{\textrm{FM}}=\eta\Omega_{0}$ is replaced by $\beta_{\textrm{AFM}}=\eta\gamma/M_{0}\left(J+2K\right)\approx\eta\gamma/M_{0}J+2K$. This difference is also manifest in the dependence of the excitation frequencies on the static magnetic field $H_{0}$: while in the ferromagnet the Larmor frequency is renormalized as $\left(1-\beta_{\textrm{FM}}\right)\gamma H_{0}$, in the antiferromagnet the corresponding factor is $\left(1+2\beta_{\textrm{AFM}}\right)^{-1}\gamma H_{0}$ for both the precession and the nutation frequencies, causing an apparent decrease in the gyromagnetic factor.

From Eqs.~\eqref{AFMprec} and \eqref{AFMnut}, the effective damping parameters in the antiferromagnet may be expressed as
\begin{align}
&\frac{{\tt Im}\left(\omega_{\textrm{p}}\right)}{{\tt Re}\left(\omega_{\textrm{p}}\right)}\approx\alpha\sqrt{\frac{\left(J+2K\right)^{2}}{4K\left(J+K\right)}}\frac{1}{\sqrt{1+2\eta\dfrac{\gamma}{M_{0}}\left(J+2K\right)}}\,, \\
&\frac{{\tt Im}\left(\omega_{\textrm{n}}\right)}{{\tt Re}\left(\omega_{\textrm{n}}\right)}\approx\alpha\frac{1+\eta\dfrac{\gamma}{M_{0}}\left(J+2K\right)}{\left[1+2\eta\dfrac{\gamma}{M_{0}}\left(J+2K\right)\right]^{\frac{3}{2}}}\,. 
\end{align}
While the inertial dynamics decrease the resonance linewidth of the antiferromagnet by a larger factor $\left(1+2\beta_{\textrm{AFM}}\right)^{-1/2}$ compared to the ferromagnet $\left(1-2\beta_{\textrm{FM}}\right)$, this is compensated by the exchange enhancement expressed in the factor $\sqrt{J/4K}$. Remarkably, the effective damping of the nutation resonance is not exchange enhanced, while it is still reduced compared to the Gilbert damping due to the inertial motion, giving rise to the particularly sharp peaks in Fig.~\ref{dissipation_antiferro}.
}

\bibliographystyle{apsrev4-2}

\begin{thebibliography}{35}%
\makeatletter
\providecommand \@ifxundefined [1]{%
 \@ifx{#1\undefined}
}%
\providecommand \@ifnum [1]{%
 \ifnum #1\expandafter \@firstoftwo
 \else \expandafter \@secondoftwo
 \fi
}%
\providecommand \@ifx [1]{%
 \ifx #1\expandafter \@firstoftwo
 \else \expandafter \@secondoftwo
 \fi
}%
\providecommand \natexlab [1]{#1}%
\providecommand \enquote  [1]{``#1''}%
\providecommand \bibnamefont  [1]{#1}%
\providecommand \bibfnamefont [1]{#1}%
\providecommand \citenamefont [1]{#1}%
\providecommand \href@noop [0]{\@secondoftwo}%
\providecommand \href [0]{\begingroup \@sanitize@url \@href}%
\providecommand \@href[1]{\@@startlink{#1}\@@href}%
\providecommand \@@href[1]{\endgroup#1\@@endlink}%
\providecommand \@sanitize@url [0]{\catcode `\\12\catcode `\$12\catcode
  `\&12\catcode `\#12\catcode `\^12\catcode `\_12\catcode `\%12\relax}%
\providecommand \@@startlink[1]{}%
\providecommand \@@endlink[0]{}%
\providecommand \url  [0]{\begingroup\@sanitize@url \@url }%
\providecommand \@url [1]{\endgroup\@href {#1}{\urlprefix }}%
\providecommand \urlprefix  [0]{URL }%
\providecommand \Eprint [0]{\href }%
\providecommand \doibase [0]{https://doi.org/}%
\providecommand \selectlanguage [0]{\@gobble}%
\providecommand \bibinfo  [0]{\@secondoftwo}%
\providecommand \bibfield  [0]{\@secondoftwo}%
\providecommand \translation [1]{[#1]}%
\providecommand \BibitemOpen [0]{}%
\providecommand \bibitemStop [0]{}%
\providecommand \bibitemNoStop [0]{.\EOS\space}%
\providecommand \EOS [0]{\spacefactor3000\relax}%
\providecommand \BibitemShut  [1]{\csname bibitem#1\endcsname}%
\let\auto@bib@innerbib\@empty
\bibitem [{\citenamefont {Stanciu}\ \emph {et~al.}(2007)\citenamefont
  {Stanciu}, \citenamefont {Hansteen}, \citenamefont {Kimel}, \citenamefont
  {Kirilyuk}, \citenamefont {Tsukamoto}, \citenamefont {Itoh},\ and\
  \citenamefont {Rasing}}]{Stanciu2007}%
  \BibitemOpen
  \bibfield  {author} {\bibinfo {author} {\bibfnamefont {C.~D.}\ \bibnamefont
  {Stanciu}}, \bibinfo {author} {\bibfnamefont {F.}~\bibnamefont {Hansteen}},
  \bibinfo {author} {\bibfnamefont {A.~V.}\ \bibnamefont {Kimel}}, \bibinfo
  {author} {\bibfnamefont {A.}~\bibnamefont {Kirilyuk}}, \bibinfo {author}
  {\bibfnamefont {A.}~\bibnamefont {Tsukamoto}}, \bibinfo {author}
  {\bibfnamefont {A.}~\bibnamefont {Itoh}},\ and\ \bibinfo {author}
  {\bibfnamefont {T.}~\bibnamefont {Rasing}},\ }\href
  {https://doi.org/10.1103/PhysRevLett.99.047601} {\bibfield  {journal}
  {\bibinfo  {journal} {Phys. Rev. Lett.}\ }\textbf {\bibinfo {volume} {99}},\
  \bibinfo {pages} {047601} (\bibinfo {year} {2007})}\BibitemShut {NoStop}%
\bibitem [{\citenamefont {Vahaplar}\ \emph {et~al.}(2009)\citenamefont
  {Vahaplar}, \citenamefont {Kalashnikova}, \citenamefont {Kimel},
  \citenamefont {Hinzke}, \citenamefont {Nowak}, \citenamefont {Chantrell},
  \citenamefont {Tsukamoto}, \citenamefont {Itoh}, \citenamefont {Kirilyuk},\
  and\ \citenamefont {Rasing}}]{Vahaplar2009}%
  \BibitemOpen
  \bibfield  {author} {\bibinfo {author} {\bibfnamefont {K.}~\bibnamefont
  {Vahaplar}}, \bibinfo {author} {\bibfnamefont {A.~M.}\ \bibnamefont
  {Kalashnikova}}, \bibinfo {author} {\bibfnamefont {A.~V.}\ \bibnamefont
  {Kimel}}, \bibinfo {author} {\bibfnamefont {D.}~\bibnamefont {Hinzke}},
  \bibinfo {author} {\bibfnamefont {U.}~\bibnamefont {Nowak}}, \bibinfo
  {author} {\bibfnamefont {R.}~\bibnamefont {Chantrell}}, \bibinfo {author}
  {\bibfnamefont {A.}~\bibnamefont {Tsukamoto}}, \bibinfo {author}
  {\bibfnamefont {A.}~\bibnamefont {Itoh}}, \bibinfo {author} {\bibfnamefont
  {A.}~\bibnamefont {Kirilyuk}},\ and\ \bibinfo {author} {\bibfnamefont
  {T.}~\bibnamefont {Rasing}},\ }\href
  {https://doi.org/10.1103/PhysRevLett.103.117201} {\bibfield  {journal}
  {\bibinfo  {journal} {Phys. Rev. Lett.}\ }\textbf {\bibinfo {volume} {103}},\
  \bibinfo {pages} {117201} (\bibinfo {year} {2009})}\BibitemShut {NoStop}%
\bibitem [{\citenamefont {Radu}\ \emph {et~al.}(2011)\citenamefont {Radu},
  \citenamefont {Vahaplar}, \citenamefont {Stamm}, \citenamefont {Kachel},
  \citenamefont {Pontius}, \citenamefont {D{\"u}rr}, \citenamefont {Ostler},
  \citenamefont {Barker}, \citenamefont {Evans}, \citenamefont {Chantrell},
  \citenamefont {Tsukamoto}, \citenamefont {Itoh}, \citenamefont {Kirilyuk},
  \citenamefont {Rasing},\ and\ \citenamefont {Kimel}}]{Radu2011}%
  \BibitemOpen
  \bibfield  {author} {\bibinfo {author} {\bibfnamefont {I.}~\bibnamefont
  {Radu}}, \bibinfo {author} {\bibfnamefont {K.}~\bibnamefont {Vahaplar}},
  \bibinfo {author} {\bibfnamefont {C.}~\bibnamefont {Stamm}}, \bibinfo
  {author} {\bibfnamefont {T.}~\bibnamefont {Kachel}}, \bibinfo {author}
  {\bibfnamefont {N.}~\bibnamefont {Pontius}}, \bibinfo {author} {\bibfnamefont
  {H.~A.}\ \bibnamefont {D{\"u}rr}}, \bibinfo {author} {\bibfnamefont {T.~A.}\
  \bibnamefont {Ostler}}, \bibinfo {author} {\bibfnamefont {J.}~\bibnamefont
  {Barker}}, \bibinfo {author} {\bibfnamefont {R.~F.~L.}\ \bibnamefont
  {Evans}}, \bibinfo {author} {\bibfnamefont {R.~W.}\ \bibnamefont
  {Chantrell}}, \bibinfo {author} {\bibfnamefont {A.}~\bibnamefont
  {Tsukamoto}}, \bibinfo {author} {\bibfnamefont {A.}~\bibnamefont {Itoh}},
  \bibinfo {author} {\bibfnamefont {A.}~\bibnamefont {Kirilyuk}}, \bibinfo
  {author} {\bibfnamefont {T.}~\bibnamefont {Rasing}},\ and\ \bibinfo {author}
  {\bibfnamefont {A.~V.}\ \bibnamefont {Kimel}},\ }\href
  {https://doi.org/10.1038/nature09901} {\bibfield  {journal} {\bibinfo
  {journal} {Nature}\ }\textbf {\bibinfo {volume} {472}},\ \bibinfo {pages}
  {205} (\bibinfo {year} {2011})}\BibitemShut {NoStop}%
\bibitem [{\citenamefont {Vahaplar}\ \emph {et~al.}(2012)\citenamefont
  {Vahaplar}, \citenamefont {Kalashnikova}, \citenamefont {Kimel},
  \citenamefont {Gerlach}, \citenamefont {Hinzke}, \citenamefont {Nowak},
  \citenamefont {Chantrell}, \citenamefont {Tsukamoto}, \citenamefont {Itoh},
  \citenamefont {Kirilyuk},\ and\ \citenamefont {Rasing}}]{vahaplar12}%
  \BibitemOpen
  \bibfield  {author} {\bibinfo {author} {\bibfnamefont {K.}~\bibnamefont
  {Vahaplar}}, \bibinfo {author} {\bibfnamefont {A.~M.}\ \bibnamefont
  {Kalashnikova}}, \bibinfo {author} {\bibfnamefont {A.~V.}\ \bibnamefont
  {Kimel}}, \bibinfo {author} {\bibfnamefont {S.}~\bibnamefont {Gerlach}},
  \bibinfo {author} {\bibfnamefont {D.}~\bibnamefont {Hinzke}}, \bibinfo
  {author} {\bibfnamefont {U.}~\bibnamefont {Nowak}}, \bibinfo {author}
  {\bibfnamefont {R.}~\bibnamefont {Chantrell}}, \bibinfo {author}
  {\bibfnamefont {A.}~\bibnamefont {Tsukamoto}}, \bibinfo {author}
  {\bibfnamefont {A.}~\bibnamefont {Itoh}}, \bibinfo {author} {\bibfnamefont
  {A.}~\bibnamefont {Kirilyuk}},\ and\ \bibinfo {author} {\bibfnamefont {{\rm
  Th}.}~\bibnamefont {Rasing}},\ }\href
  {https://doi.org/10.1103/PhysRevB.85.104402} {\bibfield  {journal} {\bibinfo
  {journal} {Phys. Rev. B}\ }\textbf {\bibinfo {volume} {85}},\ \bibinfo
  {pages} {104402} (\bibinfo {year} {2012})}\BibitemShut {NoStop}%
\bibitem [{\citenamefont {Hassdenteufel}\ \emph {et~al.}(2013)\citenamefont
  {Hassdenteufel}, \citenamefont {Hebler}, \citenamefont {Schubert},
  \citenamefont {Liebig}, \citenamefont {Teich}, \citenamefont {Helm},
  \citenamefont {Aeschlimann}, \citenamefont {Albrecht},\ and\ \citenamefont
  {Bratschitsch}}]{Hass2013}%
  \BibitemOpen
  \bibfield  {author} {\bibinfo {author} {\bibfnamefont {A.}~\bibnamefont
  {Hassdenteufel}}, \bibinfo {author} {\bibfnamefont {B.}~\bibnamefont
  {Hebler}}, \bibinfo {author} {\bibfnamefont {C.}~\bibnamefont {Schubert}},
  \bibinfo {author} {\bibfnamefont {A.}~\bibnamefont {Liebig}}, \bibinfo
  {author} {\bibfnamefont {M.}~\bibnamefont {Teich}}, \bibinfo {author}
  {\bibfnamefont {M.}~\bibnamefont {Helm}}, \bibinfo {author} {\bibfnamefont
  {M.}~\bibnamefont {Aeschlimann}}, \bibinfo {author} {\bibfnamefont
  {M.}~\bibnamefont {Albrecht}},\ and\ \bibinfo {author} {\bibfnamefont
  {R.}~\bibnamefont {Bratschitsch}},\ }\href
  {https://doi.org/https://doi.org/10.1002/adma.201300176} {\bibfield
  {journal} {\bibinfo  {journal} {Advanced Materials}\ }\textbf {\bibinfo
  {volume} {25}},\ \bibinfo {pages} {3122} (\bibinfo {year}
  {2013})}\BibitemShut {NoStop}%
\bibitem [{\citenamefont {Kimel}\ \emph {et~al.}(2009)\citenamefont {Kimel},
  \citenamefont {Ivanov}, \citenamefont {Pisarev}, \citenamefont {Usachev},
  \citenamefont {Kirilyuk},\ and\ \citenamefont {Rasing}}]{Kimel2009}%
  \BibitemOpen
  \bibfield  {author} {\bibinfo {author} {\bibfnamefont {A.~V.}\ \bibnamefont
  {Kimel}}, \bibinfo {author} {\bibfnamefont {B.~A.}\ \bibnamefont {Ivanov}},
  \bibinfo {author} {\bibfnamefont {R.~V.}\ \bibnamefont {Pisarev}}, \bibinfo
  {author} {\bibfnamefont {P.~A.}\ \bibnamefont {Usachev}}, \bibinfo {author}
  {\bibfnamefont {A.}~\bibnamefont {Kirilyuk}},\ and\ \bibinfo {author}
  {\bibfnamefont {{\rm Th}.}~\bibnamefont {Rasing}},\ }\href
  {https://doi.org/10.1038/nphys1369} {\bibfield  {journal} {\bibinfo
  {journal} {Nat. Phys.}\ }\textbf {\bibinfo {volume} {5}},\ \bibinfo {pages}
  {727} (\bibinfo {year} {2009})}\BibitemShut {NoStop}%
\bibitem [{\citenamefont {Wienholdt}\ \emph {et~al.}(2012)\citenamefont
  {Wienholdt}, \citenamefont {Hinzke},\ and\ \citenamefont
  {Nowak}}]{Wienholdt2012PRL}%
  \BibitemOpen
  \bibfield  {author} {\bibinfo {author} {\bibfnamefont {S.}~\bibnamefont
  {Wienholdt}}, \bibinfo {author} {\bibfnamefont {D.}~\bibnamefont {Hinzke}},\
  and\ \bibinfo {author} {\bibfnamefont {U.}~\bibnamefont {Nowak}},\ }\href
  {https://doi.org/10.1103/PhysRevLett.108.247207} {\bibfield  {journal}
  {\bibinfo  {journal} {Phys. Rev. Lett.}\ }\textbf {\bibinfo {volume} {108}},\
  \bibinfo {pages} {247207} (\bibinfo {year} {2012})}\BibitemShut {NoStop}%
\bibitem [{\citenamefont {Landau}\ and\ \citenamefont
  {Lifshitz}(1935)}]{landau35}%
  \BibitemOpen
  \bibfield  {author} {\bibinfo {author} {\bibfnamefont {L.~D.}\ \bibnamefont
  {Landau}}\ and\ \bibinfo {author} {\bibfnamefont {E.~M.}\ \bibnamefont
  {Lifshitz}},\ }\href@noop {} {\bibfield  {journal} {\bibinfo  {journal}
  {Phys. Z. Sowjetunion}\ }\textbf {\bibinfo {volume} {8}},\ \bibinfo {pages}
  {101} (\bibinfo {year} {1935})}\BibitemShut {NoStop}%
\bibitem [{\citenamefont {Gilbert}\ and\ \citenamefont
  {Kelly}(1955)}]{Gilbert1955}%
  \BibitemOpen
  \bibfield  {author} {\bibinfo {author} {\bibfnamefont {T.~L.}\ \bibnamefont
  {Gilbert}}\ and\ \bibinfo {author} {\bibfnamefont {J.~M.}\ \bibnamefont
  {Kelly}},\ }in\ \href@noop {} {\emph {\bibinfo {booktitle} {American
  Institute of Electrical Engineers}}}\ (\bibinfo {address} {New York},\
  \bibinfo {year} {October 1955})\ pp.\ \bibinfo {pages} {253--263}\BibitemShut
  {NoStop}%
\bibitem [{\citenamefont {{Gilbert}}(2004)}]{Gilbert2004}%
  \BibitemOpen
  \bibfield  {author} {\bibinfo {author} {\bibfnamefont {T.~L.}\ \bibnamefont
  {{Gilbert}}},\ }\href {https://doi.org/10.1109/TMAG.2004.836740} {\bibfield
  {journal} {\bibinfo  {journal} {IEEE Transactions on Magnetics}\ }\textbf
  {\bibinfo {volume} {40}},\ \bibinfo {pages} {3443} (\bibinfo {year}
  {2004})}\BibitemShut {NoStop}%
\bibitem [{\citenamefont {R\'ozsa}\ \emph {et~al.}(2019)\citenamefont
  {R\'ozsa}, \citenamefont {Selzer}, \citenamefont {Birk}, \citenamefont
  {Atxitia},\ and\ \citenamefont {Nowak}}]{Rozsa}%
  \BibitemOpen
  \bibfield  {author} {\bibinfo {author} {\bibfnamefont {L.}~\bibnamefont
  {R\'ozsa}}, \bibinfo {author} {\bibfnamefont {S.}~\bibnamefont {Selzer}},
  \bibinfo {author} {\bibfnamefont {T.}~\bibnamefont {Birk}}, \bibinfo {author}
  {\bibfnamefont {U.}~\bibnamefont {Atxitia}},\ and\ \bibinfo {author}
  {\bibfnamefont {U.}~\bibnamefont {Nowak}},\ }\href
  {https://doi.org/10.1103/PhysRevB.100.064422} {\bibfield  {journal} {\bibinfo
   {journal} {Phys. Rev. B}\ }\textbf {\bibinfo {volume} {100}},\ \bibinfo
  {pages} {064422} (\bibinfo {year} {2019})}\BibitemShut {NoStop}%
\bibitem [{\citenamefont {Gomona\u{i}}\ and\ \citenamefont
  {Loktev}(2008)}]{Gomonai}%
  \BibitemOpen
  \bibfield  {author} {\bibinfo {author} {\bibfnamefont {E.~V.}\ \bibnamefont
  {Gomona\u{i}}}\ and\ \bibinfo {author} {\bibfnamefont {V.~M.}\ \bibnamefont
  {Loktev}},\ }\href {https://doi.org/10.1063/1.2889408} {\bibfield  {journal}
  {\bibinfo  {journal} {Low Temp. Phys.}\ }\textbf {\bibinfo {volume} {34}},\
  \bibinfo {pages} {198} (\bibinfo {year} {2008})}\BibitemShut {NoStop}%
\bibitem [{\citenamefont {Hals}\ \emph {et~al.}(2011)\citenamefont {Hals},
  \citenamefont {Tserkovnyak},\ and\ \citenamefont {Brataas}}]{Hals}%
  \BibitemOpen
  \bibfield  {author} {\bibinfo {author} {\bibfnamefont {K.~M.~D.}\
  \bibnamefont {Hals}}, \bibinfo {author} {\bibfnamefont {Y.}~\bibnamefont
  {Tserkovnyak}},\ and\ \bibinfo {author} {\bibfnamefont {A.}~\bibnamefont
  {Brataas}},\ }\href {https://doi.org/10.1103/PhysRevLett.106.107206}
  {\bibfield  {journal} {\bibinfo  {journal} {Phys. Rev. Lett.}\ }\textbf
  {\bibinfo {volume} {106}},\ \bibinfo {pages} {107206} (\bibinfo {year}
  {2011})}\BibitemShut {NoStop}%
\bibitem [{\citenamefont {Gurevich}\ and\ \citenamefont
  {Melkov}(1996)}]{Gurevich}%
  \BibitemOpen
  \bibfield  {author} {\bibinfo {author} {\bibfnamefont {A.~G.}\ \bibnamefont
  {Gurevich}}\ and\ \bibinfo {author} {\bibfnamefont {G.~A.}\ \bibnamefont
  {Melkov}},\ }\href {https://books.google.se/books?id=\_ne5chyWWQIC} {\emph
  {\bibinfo {title} {Magnetization Oscillations and Waves}}},\ Lecture Notes in
  Physics\ (\bibinfo  {publisher} {CRC Press},\ \bibinfo {year}
  {1996})\BibitemShut {NoStop}%
\bibitem [{\citenamefont {Ciornei}\ \emph {et~al.}(2011)\citenamefont
  {Ciornei}, \citenamefont {Rub\'{\i}},\ and\ \citenamefont
  {Wegrowe}}]{Ciornei2011}%
  \BibitemOpen
  \bibfield  {author} {\bibinfo {author} {\bibfnamefont {M.-C.}\ \bibnamefont
  {Ciornei}}, \bibinfo {author} {\bibfnamefont {J.~M.}\ \bibnamefont
  {Rub\'{\i}}},\ and\ \bibinfo {author} {\bibfnamefont {J.-E.}\ \bibnamefont
  {Wegrowe}},\ }\href {https://doi.org/10.1103/PhysRevB.83.020410} {\bibfield
  {journal} {\bibinfo  {journal} {Phys. Rev. B}\ }\textbf {\bibinfo {volume}
  {83}},\ \bibinfo {pages} {020410} (\bibinfo {year} {2011})}\BibitemShut
  {NoStop}%
\bibitem [{\citenamefont {Wegrowe}\ and\ \citenamefont
  {Ciornei}(2012)}]{Wegrowe2012}%
  \BibitemOpen
  \bibfield  {author} {\bibinfo {author} {\bibfnamefont {J.-E.}\ \bibnamefont
  {Wegrowe}}\ and\ \bibinfo {author} {\bibfnamefont {M.-C.}\ \bibnamefont
  {Ciornei}},\ }\href {https://doi.org/10.1119/1.4709188} {\bibfield  {journal}
  {\bibinfo  {journal} {Am. J. Phys.}\ }\textbf {\bibinfo {volume} {80}},\
  \bibinfo {pages} {607} (\bibinfo {year} {2012})}\BibitemShut {NoStop}%
\bibitem [{\citenamefont {B\"ottcher}\ and\ \citenamefont
  {Henk}(2012)}]{Bottcher2012}%
  \BibitemOpen
  \bibfield  {author} {\bibinfo {author} {\bibfnamefont {D.}~\bibnamefont
  {B\"ottcher}}\ and\ \bibinfo {author} {\bibfnamefont {J.}~\bibnamefont
  {Henk}},\ }\href {https://doi.org/10.1103/PhysRevB.86.020404} {\bibfield
  {journal} {\bibinfo  {journal} {Phys. Rev. B}\ }\textbf {\bibinfo {volume}
  {86}},\ \bibinfo {pages} {020404} (\bibinfo {year} {2012})}\BibitemShut
  {NoStop}%
\bibitem [{\citenamefont {F\"ahnle}\ \emph {et~al.}(2011)\citenamefont
  {F\"ahnle}, \citenamefont {Steiauf},\ and\ \citenamefont
  {Illg}}]{Fahnle2011}%
  \BibitemOpen
  \bibfield  {author} {\bibinfo {author} {\bibfnamefont {M.}~\bibnamefont
  {F\"ahnle}}, \bibinfo {author} {\bibfnamefont {D.}~\bibnamefont {Steiauf}},\
  and\ \bibinfo {author} {\bibfnamefont {C.}~\bibnamefont {Illg}},\ }\href
  {https://doi.org/10.1103/PhysRevB.84.172403} {\bibfield  {journal} {\bibinfo
  {journal} {Phys. Rev. B}\ }\textbf {\bibinfo {volume} {84}},\ \bibinfo
  {pages} {172403} (\bibinfo {year} {2011})}\BibitemShut {NoStop}%
\bibitem [{\citenamefont {F\"ahnle}\ and\ \citenamefont
  {Illg}(2011)}]{Fahnle2011JPCM}%
  \BibitemOpen
  \bibfield  {author} {\bibinfo {author} {\bibfnamefont {M.}~\bibnamefont
  {F\"ahnle}}\ and\ \bibinfo {author} {\bibfnamefont {C.}~\bibnamefont
  {Illg}},\ }\href {http://stacks.iop.org/0953-8984/23/i=49/a=493201}
  {\bibfield  {journal} {\bibinfo  {journal} {J. Phys.: Condens. Matter}\
  }\textbf {\bibinfo {volume} {23}},\ \bibinfo {pages} {493201} (\bibinfo
  {year} {2011})}\BibitemShut {NoStop}%
\bibitem [{\citenamefont {Bhattacharjee}\ \emph {et~al.}(2012)\citenamefont
  {Bhattacharjee}, \citenamefont {Nordstr\"om},\ and\ \citenamefont
  {Fransson}}]{Bhattacharjee2012}%
  \BibitemOpen
  \bibfield  {author} {\bibinfo {author} {\bibfnamefont {S.}~\bibnamefont
  {Bhattacharjee}}, \bibinfo {author} {\bibfnamefont {L.}~\bibnamefont
  {Nordstr\"om}},\ and\ \bibinfo {author} {\bibfnamefont {J.}~\bibnamefont
  {Fransson}},\ }\href {https://doi.org/10.1103/PhysRevLett.108.057204}
  {\bibfield  {journal} {\bibinfo  {journal} {Phys. Rev. Lett.}\ }\textbf
  {\bibinfo {volume} {108}},\ \bibinfo {pages} {057204} (\bibinfo {year}
  {2012})}\BibitemShut {NoStop}%
\bibitem [{\citenamefont {Mondal}\ \emph {et~al.}(2017)\citenamefont {Mondal},
  \citenamefont {Berritta}, \citenamefont {Nandy},\ and\ \citenamefont
  {Oppeneer}}]{Mondal2017Nutation}%
  \BibitemOpen
  \bibfield  {author} {\bibinfo {author} {\bibfnamefont {R.}~\bibnamefont
  {Mondal}}, \bibinfo {author} {\bibfnamefont {M.}~\bibnamefont {Berritta}},
  \bibinfo {author} {\bibfnamefont {A.~K.}\ \bibnamefont {Nandy}},\ and\
  \bibinfo {author} {\bibfnamefont {P.~M.}\ \bibnamefont {Oppeneer}},\ }\href
  {https://doi.org/10.1103/PhysRevB.96.024425} {\bibfield  {journal} {\bibinfo
  {journal} {Phys. Rev. B}\ }\textbf {\bibinfo {volume} {96}},\ \bibinfo
  {pages} {024425} (\bibinfo {year} {2017})}\BibitemShut {NoStop}%
\bibitem [{\citenamefont {Mondal}\ \emph {et~al.}(2018)\citenamefont {Mondal},
  \citenamefont {Berritta},\ and\ \citenamefont {Oppeneer}}]{Mondal2018JPCM}%
  \BibitemOpen
  \bibfield  {author} {\bibinfo {author} {\bibfnamefont {R.}~\bibnamefont
  {Mondal}}, \bibinfo {author} {\bibfnamefont {M.}~\bibnamefont {Berritta}},\
  and\ \bibinfo {author} {\bibfnamefont {P.~M.}\ \bibnamefont {Oppeneer}},\
  }\href {http://stacks.iop.org/0953-8984/30/i=26/a=265801} {\bibfield
  {journal} {\bibinfo  {journal} {J. Phys.: Condens. Matter}\ }\textbf
  {\bibinfo {volume} {30}},\ \bibinfo {pages} {265801} (\bibinfo {year}
  {2018})}\BibitemShut {NoStop}%
\bibitem [{\citenamefont {Li}\ \emph {et~al.}(2015)\citenamefont {Li},
  \citenamefont {Barra}, \citenamefont {Auffret}, \citenamefont {Ebels},\ and\
  \citenamefont {Bailey}}]{Li2015}%
  \BibitemOpen
  \bibfield  {author} {\bibinfo {author} {\bibfnamefont {Y.}~\bibnamefont
  {Li}}, \bibinfo {author} {\bibfnamefont {A.-L.}\ \bibnamefont {Barra}},
  \bibinfo {author} {\bibfnamefont {S.}~\bibnamefont {Auffret}}, \bibinfo
  {author} {\bibfnamefont {U.}~\bibnamefont {Ebels}},\ and\ \bibinfo {author}
  {\bibfnamefont {W.~E.}\ \bibnamefont {Bailey}},\ }\href
  {https://doi.org/10.1103/PhysRevB.92.140413} {\bibfield  {journal} {\bibinfo
  {journal} {Phys. Rev. B}\ }\textbf {\bibinfo {volume} {92}},\ \bibinfo
  {pages} {140413} (\bibinfo {year} {2015})}\BibitemShut {NoStop}%
\bibitem [{\citenamefont {Thonig}\ \emph {et~al.}(2017)\citenamefont {Thonig},
  \citenamefont {Eriksson},\ and\ \citenamefont {Pereiro}}]{Thonig2017}%
  \BibitemOpen
  \bibfield  {author} {\bibinfo {author} {\bibfnamefont {D.}~\bibnamefont
  {Thonig}}, \bibinfo {author} {\bibfnamefont {O.}~\bibnamefont {Eriksson}},\
  and\ \bibinfo {author} {\bibfnamefont {M.}~\bibnamefont {Pereiro}},\ }\href
  {https://doi.org/10.1038/s41598-017-01081-z} {\bibfield  {journal} {\bibinfo
  {journal} {Sci. Rep.}\ }\textbf {\bibinfo {volume} {7}},\ \bibinfo {pages}
  {931} (\bibinfo {year} {2017})}\BibitemShut {NoStop}%
\bibitem [{\citenamefont {Neeraj}\ \emph {et~al.}(2020)\citenamefont {Neeraj},
  \citenamefont {Awari}, \citenamefont {Kovalev}, \citenamefont {Polley},
  \citenamefont {Zhou~Hagstr{\"o}m}, \citenamefont {Arekapudi}, \citenamefont
  {Semisalova}, \citenamefont {Lenz}, \citenamefont {Green}, \citenamefont
  {Deinert}, \citenamefont {Ilyakov}, \citenamefont {Chen}, \citenamefont
  {Bawatna}, \citenamefont {Scalera}, \citenamefont {d'Aquino}, \citenamefont
  {Serpico}, \citenamefont {Hellwig}, \citenamefont {Wegrowe}, \citenamefont
  {Gensch},\ and\ \citenamefont {Bonetti}}]{neeraj2019experimental}%
  \BibitemOpen
  \bibfield  {author} {\bibinfo {author} {\bibfnamefont {K.}~\bibnamefont
  {Neeraj}}, \bibinfo {author} {\bibfnamefont {N.}~\bibnamefont {Awari}},
  \bibinfo {author} {\bibfnamefont {S.}~\bibnamefont {Kovalev}}, \bibinfo
  {author} {\bibfnamefont {D.}~\bibnamefont {Polley}}, \bibinfo {author}
  {\bibfnamefont {N.}~\bibnamefont {Zhou~Hagstr{\"o}m}}, \bibinfo {author}
  {\bibfnamefont {S.~S. P.~K.}\ \bibnamefont {Arekapudi}}, \bibinfo {author}
  {\bibfnamefont {A.}~\bibnamefont {Semisalova}}, \bibinfo {author}
  {\bibfnamefont {K.}~\bibnamefont {Lenz}}, \bibinfo {author} {\bibfnamefont
  {B.}~\bibnamefont {Green}}, \bibinfo {author} {\bibfnamefont {J.-C.}\
  \bibnamefont {Deinert}}, \bibinfo {author} {\bibfnamefont {I.}~\bibnamefont
  {Ilyakov}}, \bibinfo {author} {\bibfnamefont {M.}~\bibnamefont {Chen}},
  \bibinfo {author} {\bibfnamefont {M.}~\bibnamefont {Bawatna}}, \bibinfo
  {author} {\bibfnamefont {V.}~\bibnamefont {Scalera}}, \bibinfo {author}
  {\bibfnamefont {M.}~\bibnamefont {d'Aquino}}, \bibinfo {author}
  {\bibfnamefont {C.}~\bibnamefont {Serpico}}, \bibinfo {author} {\bibfnamefont
  {O.}~\bibnamefont {Hellwig}}, \bibinfo {author} {\bibfnamefont {J.-E.}\
  \bibnamefont {Wegrowe}}, \bibinfo {author} {\bibfnamefont {M.}~\bibnamefont
  {Gensch}},\ and\ \bibinfo {author} {\bibfnamefont {S.}~\bibnamefont
  {Bonetti}},\ }\href {https://doi.org/10.1038/s41567-020-01040-y} {\bibfield
  {journal} {\bibinfo  {journal} {Nat. Phys.}\ } (\bibinfo {year} {2020})},\
  \Eprint {https://arxiv.org/abs/https://doi.org/10.1038/s41567-020-01040-y}
  {https://doi.org/10.1038/s41567-020-01040-y} \BibitemShut {NoStop}%
\bibitem [{\citenamefont {Olive}\ \emph {et~al.}(2012)\citenamefont {Olive},
  \citenamefont {Lansac},\ and\ \citenamefont {Wegrowe}}]{Olive2012}%
  \BibitemOpen
  \bibfield  {author} {\bibinfo {author} {\bibfnamefont {E.}~\bibnamefont
  {Olive}}, \bibinfo {author} {\bibfnamefont {Y.}~\bibnamefont {Lansac}},\ and\
  \bibinfo {author} {\bibfnamefont {J.-E.}\ \bibnamefont {Wegrowe}},\ }\href
  {https://doi.org/10.1063/1.4712056} {\bibfield  {journal} {\bibinfo
  {journal} {Appl. Phys. Lett.}\ }\textbf {\bibinfo {volume} {100}},\ \bibinfo
  {pages} {192407} (\bibinfo {year} {2012})}\BibitemShut {NoStop}%
\bibitem [{\citenamefont {Olive}\ \emph {et~al.}(2015)\citenamefont {Olive},
  \citenamefont {Lansac}, \citenamefont {Meyer}, \citenamefont {Hayoun},\ and\
  \citenamefont {Wegrowe}}]{Olive2015}%
  \BibitemOpen
  \bibfield  {author} {\bibinfo {author} {\bibfnamefont {E.}~\bibnamefont
  {Olive}}, \bibinfo {author} {\bibfnamefont {Y.}~\bibnamefont {Lansac}},
  \bibinfo {author} {\bibfnamefont {M.}~\bibnamefont {Meyer}}, \bibinfo
  {author} {\bibfnamefont {M.}~\bibnamefont {Hayoun}},\ and\ \bibinfo {author}
  {\bibfnamefont {J.-E.}\ \bibnamefont {Wegrowe}},\ }\href
  {https://doi.org/10.1063/1.4921908} {\bibfield  {journal} {\bibinfo
  {journal} {J. Appl. Phys.}\ }\textbf {\bibinfo {volume} {117}},\ \bibinfo
  {pages} {213904} (\bibinfo {year} {2015})}\BibitemShut {NoStop}%
\bibitem [{\citenamefont {Cherkasskii}\ \emph {et~al.}(2020)\citenamefont
  {Cherkasskii}, \citenamefont {Farle},\ and\ \citenamefont
  {Semisalova}}]{cherkasskii2020nutation}%
  \BibitemOpen
  \bibfield  {author} {\bibinfo {author} {\bibfnamefont {M.}~\bibnamefont
  {Cherkasskii}}, \bibinfo {author} {\bibfnamefont {M.}~\bibnamefont {Farle}},\
  and\ \bibinfo {author} {\bibfnamefont {A.}~\bibnamefont {Semisalova}},\
  }\href@noop {} {\bibinfo {title} {Nutation resonance in ferromagnets}}
  (\bibinfo {year} {2020}),\ \Eprint {https://arxiv.org/abs/2008.12221}
  {arXiv:2008.12221 [cond-mat.mes-hall]} \BibitemShut {NoStop}%
\bibitem [{\citenamefont {Makhfudz}\ \emph {et~al.}(2020)\citenamefont
  {Makhfudz}, \citenamefont {Olive},\ and\ \citenamefont
  {Nicolis}}]{Makhfudz2020}%
  \BibitemOpen
  \bibfield  {author} {\bibinfo {author} {\bibfnamefont {I.}~\bibnamefont
  {Makhfudz}}, \bibinfo {author} {\bibfnamefont {E.}~\bibnamefont {Olive}},\
  and\ \bibinfo {author} {\bibfnamefont {S.}~\bibnamefont {Nicolis}},\ }\href
  {https://doi.org/10.1063/5.0013062} {\bibfield  {journal} {\bibinfo
  {journal} {Appl. Phys. Lett.}\ }\textbf {\bibinfo {volume} {117}},\ \bibinfo
  {pages} {132403} (\bibinfo {year} {2020})}\BibitemShut {NoStop}%
\bibitem [{\citenamefont {Kikuchi}\ and\ \citenamefont
  {Tatara}(2015)}]{Kikuchi}%
  \BibitemOpen
  \bibfield  {author} {\bibinfo {author} {\bibfnamefont {T.}~\bibnamefont
  {Kikuchi}}\ and\ \bibinfo {author} {\bibfnamefont {G.}~\bibnamefont
  {Tatara}},\ }\href {https://doi.org/10.1103/PhysRevB.92.184410} {\bibfield
  {journal} {\bibinfo  {journal} {Phys. Rev. B}\ }\textbf {\bibinfo {volume}
  {92}},\ \bibinfo {pages} {184410} (\bibinfo {year} {2015})}\BibitemShut
  {NoStop}%
\bibitem [{\citenamefont {Kittel}(1951)}]{Kittel1951}%
  \BibitemOpen
  \bibfield  {author} {\bibinfo {author} {\bibfnamefont {C.}~\bibnamefont
  {Kittel}},\ }\href {https://doi.org/10.1103/PhysRev.82.565} {\bibfield
  {journal} {\bibinfo  {journal} {Phys. Rev.}\ }\textbf {\bibinfo {volume}
  {82}},\ \bibinfo {pages} {565} (\bibinfo {year} {1951})}\BibitemShut
  {NoStop}%
\bibitem [{\citenamefont {Keffer}\ and\ \citenamefont {Kittel}(1952)}]{Keffer}%
  \BibitemOpen
  \bibfield  {author} {\bibinfo {author} {\bibfnamefont {F.}~\bibnamefont
  {Keffer}}\ and\ \bibinfo {author} {\bibfnamefont {C.}~\bibnamefont
  {Kittel}},\ }\href {https://doi.org/10.1103/PhysRev.85.329} {\bibfield
  {journal} {\bibinfo  {journal} {Phys. Rev.}\ }\textbf {\bibinfo {volume}
  {85}},\ \bibinfo {pages} {329} (\bibinfo {year} {1952})}\BibitemShut
  {NoStop}%
\bibitem [{\citenamefont {Kamra}\ \emph {et~al.}(2018)\citenamefont {Kamra},
  \citenamefont {Troncoso}, \citenamefont {Belzig},\ and\ \citenamefont
  {Brataas}}]{Kamra2018}%
  \BibitemOpen
  \bibfield  {author} {\bibinfo {author} {\bibfnamefont {A.}~\bibnamefont
  {Kamra}}, \bibinfo {author} {\bibfnamefont {R.~E.}\ \bibnamefont {Troncoso}},
  \bibinfo {author} {\bibfnamefont {W.}~\bibnamefont {Belzig}},\ and\ \bibinfo
  {author} {\bibfnamefont {A.}~\bibnamefont {Brataas}},\ }\href
  {https://doi.org/10.1103/PhysRevB.98.184402} {\bibfield  {journal} {\bibinfo
  {journal} {Phys. Rev. B}\ }\textbf {\bibinfo {volume} {98}},\ \bibinfo
  {pages} {184402} (\bibinfo {year} {2018})}\BibitemShut {NoStop}%
\bibitem [{\citenamefont {Schlickeiser}\ \emph {et~al.}(2012)\citenamefont
  {Schlickeiser}, \citenamefont {Atxitia}, \citenamefont {Wienholdt},
  \citenamefont {Hinzke}, \citenamefont {Chubykalo-Fesenko},\ and\
  \citenamefont {Nowak}}]{Frank2012}%
  \BibitemOpen
  \bibfield  {author} {\bibinfo {author} {\bibfnamefont {F.}~\bibnamefont
  {Schlickeiser}}, \bibinfo {author} {\bibfnamefont {U.}~\bibnamefont
  {Atxitia}}, \bibinfo {author} {\bibfnamefont {S.}~\bibnamefont {Wienholdt}},
  \bibinfo {author} {\bibfnamefont {D.}~\bibnamefont {Hinzke}}, \bibinfo
  {author} {\bibfnamefont {O.}~\bibnamefont {Chubykalo-Fesenko}},\ and\
  \bibinfo {author} {\bibfnamefont {U.}~\bibnamefont {Nowak}},\ }\href
  {https://doi.org/10.1103/PhysRevB.86.214416} {\bibfield  {journal} {\bibinfo
  {journal} {Phys. Rev. B}\ }\textbf {\bibinfo {volume} {86}},\ \bibinfo
  {pages} {214416} (\bibinfo {year} {2012})}\BibitemShut {NoStop}%
  \bibitem [{\citenamefont {Hutchings}\ and\ \citenamefont
  {Samuelsen}(1972)}]{Hutchings1972}%
  \BibitemOpen
  \bibfield  {author} {\bibinfo {author} {\bibfnamefont {M.~T.}\ \bibnamefont
  {Hutchings}}\ and\ \bibinfo {author} {\bibfnamefont {E.~J.}\ \bibnamefont
  {Samuelsen}},\ }\href {\doibase 10.1103/PhysRevB.6.3447} {\bibfield
  {journal} {\bibinfo  {journal} {Phys. Rev. B}\ }\textbf {\bibinfo {volume}
  {6}},\ \bibinfo {pages} {3447} (\bibinfo {year} {1972})}\BibitemShut
  {NoStop}%
  \bibitem [{\citenamefont {Besnus}\ and\ \citenamefont
  {Meyer}(1973)}]{Besnus1973}%
  \BibitemOpen
  \bibfield  {author} {\bibinfo {author} {\bibfnamefont {M.~J.}\ \bibnamefont
  {Besnus}}\ and\ \bibinfo {author} {\bibfnamefont {A.~J.~P.}\ \bibnamefont
  {Meyer}},\ }\href {\doibase 10.1002/pssb.2220580213} {\bibfield  {journal}
  {\bibinfo  {journal} {phys. stat. sol. (b)}\ }\textbf {\bibinfo {volume}
  {58}},\ \bibinfo {pages} {533} (\bibinfo {year} {1973})}\BibitemShut
  {NoStop}%
\bibitem [{\citenamefont {Zhang}\ \emph {et~al.}(2012)\citenamefont {Zhang},
  \citenamefont {Skomski}, \citenamefont {Li}, \citenamefont {Li},
  \citenamefont {Manchanda}, \citenamefont {Kashyap}, \citenamefont {Kirby},
  \citenamefont {Liou},\ and\ \citenamefont {Sellmyer}}]{Zhang2012JAP}%
  \BibitemOpen
  \bibfield  {author} {\bibinfo {author} {\bibfnamefont {R.}~\bibnamefont
  {Zhang}}, \bibinfo {author} {\bibfnamefont {R.}~\bibnamefont {Skomski}},
  \bibinfo {author} {\bibfnamefont {X.}~\bibnamefont {Li}}, \bibinfo {author}
  {\bibfnamefont {Z.}~\bibnamefont {Li}}, \bibinfo {author} {\bibfnamefont
  {P.}~\bibnamefont {Manchanda}}, \bibinfo {author} {\bibfnamefont
  {A.}~\bibnamefont {Kashyap}}, \bibinfo {author} {\bibfnamefont {R.~D.}\
  \bibnamefont {Kirby}}, \bibinfo {author} {\bibfnamefont {S.-H.}\ \bibnamefont
  {Liou}}, \ and\ \bibinfo {author} {\bibfnamefont {D.~J.}\ \bibnamefont
  {Sellmyer}},\ }\href {\doibase 10.1063/1.3677928} {\bibfield  {journal}
  {\bibinfo  {journal} {J. Appl. Phys.}\ }\textbf {\bibinfo {volume} {111}},\
  \bibinfo {pages} {07D720} (\bibinfo {year} {2012})}\BibitemShut {NoStop}%
  \bibitem [{\citenamefont {Nowak}(2007)}]{nowak2007handbook}%
  \BibitemOpen
  \bibfield  {author} {\bibinfo {author} {\bibfnamefont {U.}~\bibnamefont
  {Nowak}},\ }\href@noop {} {\bibinfo {title} {Handbook of {M}agnetism and
  {A}dvanced {M}agnetic {M}aterials}} (\bibinfo {year} {2007})\BibitemShut
  {NoStop}%
\end{thebibliography}

\end{document}